\documentclass[usenatbib]{mnras}

\usepackage[T1]{fontenc}
\usepackage{ae,aecompl}
\usepackage{soul}

\usepackage{graphicx}	
\usepackage{amsmath}	
\usepackage{amssymb}	

\newcommand{\msun}{\text{M}_\odot}
\newcommand{\Zsun}{Z_\odot}

\title[PISNe rates and host galaxy properties]{The rates and host galaxies of pair-instability supernovae through cosmic time: Predictions from BPASS and IllustrisTNG}

\author[Briel, M. \& Metha, B. et al.]{
Max M. Briel$^{1,2,3}$\thanks{These authors contributed equally to this work.}\thanks{max.briel@gmail.com},
Benjamin Metha$^{4,5}$\footnotemark[1]\thanks{methab@student.unimelb.edu.au}, 
Jan J. Eldridge$^{2}$, Takashi J. Moriya$^{6,7,8}$,
Michele Trenti$^{4,5}$
\\
$^1$Departement d’Astronomie, Université de Genève, Chemin Pegasi 51, CH-1290 Versoix, Switzerland\\
$^2$Department of Physics, University of Auckland, Private Bag 92019, Auckland, New Zealand\\
$^3$Gravitational Wave Science Center (GWSC), Université de Genève, CH1211 Geneva, Switzerland\\
$^4$School of Physics, The University of Melbourne, VIC 3010, Australia\\
$^5$Australian Research Council Centre of Excellence for All-Sky Astrophysics in 3-Dimensions, Australia\\
$^6$National Astronomical Observatory of Japan, National Institutes of Natural Sciences, 2-21-1 Osawa, Mitaka, Tokyo 181-8588, Japan\\
$^7$Graduate Institute for Advanced Studies, SOKENDAI, 2-21-1 Osawa, Mitaka, Tokyo 181-8588, Japan \\
$^8$School of Physics and Astronomy, Faculty of Science, Monash University, Clayton, VIC 3800, Australia
}

\date{Accepted 2024 August 23. Received 2024 August 23; in original form 2024 May 31}

\pubyear{2024}

\begin{document}
\label{firstpage}
\pagerange{\pageref{firstpage}--\pageref{lastpage}}
\maketitle

\begin{abstract}
Pair-instability supernovae (PISNe) have long been predicted to be the final fates of near-zero-metallicity very massive stars ($Z < Z_\odot/3$, M$_\mathrm{ZAMS} \gtrsim 140 \msun$). However, no definite PISN has been observed to date, leaving theoretical modelling validation open. To investigate the observability of these explosive transients, we combine detailed stellar evolution models for PISNe formation, computed from the Binary Population and Spectral Synthesis code suite, \textsc{bpass}, with the star formation history of all individual computational elements in the Illustris-TNG simulation. This allows us to compute comic PISN rates and predict their host galaxy properties. Of particular importance is that IllustrisTNG galaxies do not have uniform metallicities throughout, with metal-enriched galaxies often harbouring metal-poor pockets of gas where PISN progenitors may form.
Accounting for the chemical inhomogeneities within these galaxies, we find that the peak redshift of PISNe formation is $z=3.5$ instead of the value of $z=6$ when ignoring chemical inhomogeneities within galaxies. Furthermore, the rate increases by an order of magnitude from 1.9 to 29 PISN Gpc$^{-3}$ yr$^{-1}$ at $z=0$, if the chemical inhomogeneities are considered.
Using state-of-the-art theoretical PISN light curves, we find an observed rate of $13.8$ (1.2) visible PISNe per year for the Euclid-Deep survey, or $83$ (7.3) over the six-year lifetime of the mission when considering chemically inhomogeneous (homogenous) systems. Interestingly, only 12 per cent of helium PISN progenitors are sufficiently massive to power a super-luminous supernova event, which can potentially explain why PISN identification in time-domain surveys remains elusive and progress requires dedicated strategies.
\end{abstract}

\begin{keywords}
transients: supernovae -- galaxies: abundances -- galaxies: evolution -- binaries: general
\end{keywords} 


\section{Introduction} \label{sec:intro}

Evolved very massive and low-metallicity stars (with zero-age-main-sequence mass $M_\mathrm{ZAMS} \gtrsim 140 \msun$, and metallicity $Z/Z_{\odot}<1/3$) are predicted to end their lives in a pair-instability supernova (PISN), where the star is disrupted completely, leaving no remnant \citep[e.g.][]{Fowler+Holye64, Barkat+67}. These events eject a large amount of material and energy into the environment, enriching galaxies with more massive elements \citep{langer_2012}. This process also limits the formation of black holes (BH) above ${\sim}40 \msun$ \citep[e.g.][]{Heger+Woosley02, farmer_2019, Woosley+Heger21}.
Theoretical PISN rate estimates are similar to the observed super-luminous supernova rate \citep{Briel+21, Moriya+22-Euclid, Tanikawa+23, hendriks_2023}. However, no smoking-gun PISN have been observed to date \citep[although see][]{Schulze+23}.

PISNe occur inside the cores of evolved very massive stars due to high temperatures and relatively low densities enabling electron-positron pair production. This process removes the outwards radiative photon pressure inducing a collapse of the star. If the star is massive enough ($\mathrm{M}_\mathrm{ZAMS} \gtrsim 140 \msun$), the explosive carbon-oxygen burning that follows completely unbinds the star in a PISN \citep[e.g.][]{Fowler+Holye64, Zeldovich+71}.
For single stars with ZAMS masses above $140 \msun$ to experience a PISN, they require a helium core above ${\sim}65 \msun$ at the time of core-collapse \citep{Woosley+17}, which limits PISNe to lower metallicity environments, depending on the implemented stellar wind mass loss prescriptions \citep{langer_2007}. A helium core above ${\sim}130 \msun$ will instead lead to a direct collapse of the star into a BH  through the photodisintegration of heavy elements \citep{woosley_2002}.
While the PISN explosion mechanism is robust  \citep{El_Eid+Langer86, Heger+Woosley02, Waldman08}, the mass limits are dependent on many factors \citep{Woosley+Heger21}, such as the nuclear reaction rates \citep[e.g.][]{Takahashi18, farmer_2019} and stellar rotation \citep[e.g.][]{Fowler+Holye64, Marchant+Moriya20, Umeda+24}.

Since stars with helium cores between ${\sim}65 \msun$ and ${\sim}130$ will not collapse into a BH, PISN limits the formation of BHs between ${\sim}40 \msun$ and ${\sim}120 \msun$ \citep[e.g.][]{Heger+Woosley02, farmer_2019, Woosley+Heger21}, thus shaping the mass distribution of merging binary BHs \citep[e.g.][]{Marchant+16, farmer_2019, duBuisson+20}. However, no clear evidence for such a dearth in the binary BH mass distribution has been found by the LIGO/Virgo/KAGRA collaboration \citep{abbott+2023, Rinaldi+24}. This could indicate a higher mass limit of PISN \citep[e.g.][]{stevenson_2019}, or other different formation mechanism for merging binary BHs in PISN mass gap, such as hierarchical mergers \citep[e.g.][]{Belczynski+20, gerosa_21, Mapelli+21, Anagnostou+22, Rose+22}, Population-III stars \citep{Hijikawa+21, Santoliquido+23, Iwaya+23, Tanikawa+24}, or stable mass transfer with super-Eddington accretion \citep{Briel+23}.

PISNe eject large amounts of material in their surrounding, and a single PISN could release tens of solar masses of chemical enrichment into its surrounding environment \citep{langer_2012}. The next-generation stars formed from this ejected material will have a unique chemical signature due to the nucleosynthesis processes in the PISN \citep[see][]{Heger+Woosley02, nomoto_2002, nomoto_2013}. Several very metal-poor stars with such signatures have been observed \citep[e.g.][]{aoki+14, Salvadori+19, xing_2023}, providing indirect evidence for the occurrence of PISNe with similar results from metal-poor galaxies \citep[e.g.][]{Isobe+22}.

The strongest evidence for the existence of PISN would be a direct observation of the thermonuclear explosion. Theoretical PISN light curve models predict a luminous and long-lived transient, similar to a super-luminous supernova (SLSN) but are powered by $^{56}$Ni instead of a central engine \citep{Kasen+11, Kozyreva+17}.
Several SLSNe have exhibited features predicted by PISN models \citep[see ][for an overview]{Schulze+23}, but none have completely matched all expected PISN features \citep{Schulze+23} with a central engine being the favoured scenario to explain hydrogen-poor (Type-I) SLSNe. Only a handful of potential Type-I SLSNe are not well explained by this scenario \citep{Umeda+24}. The strongest of these is SN2018ibb \citep{Schulze+23}, where the central engine model is disfavoured due to the absence of light curve flattening, and the PISN model matches nearly all observed features. However, the high blackbody temperature and long rise time of SN2018ibb make this event difficult to match to theoretical PISN light curves \citep{nagele_2024}.

Although PISNe are expected to be bright, they are thought to be limited to high-redshift galaxies because of their metallicity threshold, hampering opportunities to directly detect them.
Metal-free gas could provide the ideal birth sites for Population III stars, which are expected to be a major contributor to the PISN rate at $z>6$ due to the limited stellar mass loss from winds \citep[e.g.][]{Tanikawa+23}.
On the other hand, the star formation rate of Population III stars could continue up to low redshifts depending on the assumed simulation physics, as high-resolution cosmological simulations have shown \citep{wise_2012, xu_2016, liu_2020c}.
These metal-free pockets after the epoch of reionization could provide birth sites for Population III PISN progenitors up to present day \citep{tornatore_2007, trenti_2009, venditti_2023}.

One thing is clear; PISN progenitor stars are required to have low metallicities, such that their stellar winds are weak, and the PISN progenitor can retain a large helium core \citep{Heger+Woosley02, woosley_2002}.
As such, PISN rate predictions primarily focus on the contribution of Population III or very metal-poor Population II stars \citep{langer_2007, spera_2017, venditti_2023, wiggins_2024}. This rate ranges from $10^{-2}$ to $10^2$ event per year per deg$^2$ and is dominated by the uncertainty in the early star formation and initial mass functions.

Furthermore, detailed binary evolution grids have shown that mergers can result in PISN progenitors at SMC \citep[${\sim}0.25 Z_\odot$; see for example][]{choudhury_2018} metallicity \citep{vigna-gomez_2019}. Together with chemically homogeneous evolution in close binaries \citep{marchant_2019}, the PISN rate could be boosted when considering a larger metallicity range.
Including Population I/II single stars and binaries, the PISN can range between 1 and 80 events per year per Gpc$^3$ depending on the PISN mass range \citep{hendriks_2023} or star formation history \citep{Briel+21}.
\citet{Tanikawa+23} also explored the impact of a higher PISN mass range and an extended IMF to $600 \msun$, while including a Population III contribution. They found that the Euclid Deep Survey might be able to detect $10^{-3}$ to 4 PISN over 6 years of observations.

In these binary population synthesis studies, the rate of PISNe throughout cosmic time is predicted either (i) assuming that the metallicity of all stars is at the mean metallicity of the Cosmos at that redshift, or (ii) accounting for differences in metallicities of different galaxies, but assuming that all stars formed in each galaxy have the same metallicity.
However, galaxies are known to be chemically inhomogeneous \citep[e.g.][]{Aller1942, Searle71, VilaCostas+92, Zaritsky+94, CALIFA, Ho+15, SAMI, MANGA, lenstronometals}. While slit spectroscopy generally returns only one metallicity measurement for a high-redshift galaxy, it is incorrect to extrapolate this data to conclude that all gas within the galaxy has the same metallicity. Simulations have shown that galaxies have a variety of metallicities within them, even up to high redshifts ($z \sim 3$; e.g. \citealt{Hemler+21, Yates+21, ZZ2, Garcia+23, Garcia+24}).
Because of this, galaxies with a high average metallicity may harbour low-metallicity ``pockets" of gas where PISN progenitors may be born, similar to GRBs \citep{Perley+16, Niino+17, Metha+20, ZZ1}. 

In this paper, we investigate how accounting for the inhomogeneous chemical structure of galaxies affects both the rate of PISNe from Population I/II stars as a function of time, and the predicted host galaxy properties, by combining predicted galaxy population properties from the IllustrisTNG-50 simulation with the detailed stellar models from the Binary Population and Spectral Synthesis (\textsc{bpass}) code suite. The \textsc{bpass} models are used to predict the PISN formation efficiency as a function of the metallicity of the progenitor star, or binary progenitor system. In Section \ref{sec:methods}, we describe the IllustrisTNG-50 simulation and the PISN metallicity bias functions as computed using the \textsc{bpass} models. We make predictions for the rates of PISNe throughout cosmic time in Section \ref{sec:results}, and use this result to make predictions about the number of PISNe that are expected to be seen both in archival data and by the forthcoming Euclid Deep Survey in Section \ref{sec:visible_rates}. In Section \ref{sec:host-gals}, we make predictions for the host galaxies of PISNe throughout cosmic time. We discuss our findings in Section \ref{sec:discussion}, focusing on the implications of the potential non-detection of PISNe on stellar physics. We summarise our main results in Section \ref{sec:conclusions}.

Throughout this paper, we follow the conventions of the IllustrisTNG simulation and assume a Flat $\Lambda$CDM cosmology using the cosmological parameters of \citet{Planck15}. We take the value of solar metallicity to be equal to $Z_\odot = 0.0139$ \citep{Asplund+21}.

\section{Methods} \label{sec:methods}

\subsection{Population synthesis and pair-instability supernova selection}
For population synthesis, we use \textsc{bpass} v2.2 \citep{Eldridge+17, Stanway+Eldridge18}, which contains detailed stellar models evolved using an adapted version of the Cambridge STARS code \citep{Eggleton71}.
Initially, the primary star is evolved in detail using the 1D stellar evolution code, while the evolution of the secondary is approximated using rapid single star equations from \citet{hurley_2000}. Once the primary becomes a white dwarf (WD) or undergoes a supernova or PISN, the secondary is evolved in detail \citep[for a more comprehensive explanation of the \textsc{bpass} models, see][]{Eldridge+17, Stanway+Eldridge18, stevance_2022, Briel+23}. If a companion accretes more than 5 per cent of its initial mass at metallicities below $Z=0.43Z_\odot$, \textsc{bpass} implements a quasi-chemically homogeneous evolution (QHE). Furthermore, a merger occurs when both stars fill their Roche lobe and their sum is larger than the separation of the system. The companion mass is added to the primary models using the surface abundances of the primary \citep{Eldridge+17, Stanway+Eldridge18}. We use the \citet{Chabrier03} initial mass function extended up to $300 \msun$, and initialise the binary properties according to empirical relations by \citet{moe_2018}.

\textsc{bpass} spans 13 metallicities from $Z=0.0007Z_\odot$ to $Z=2.9Z_\odot$\footnote{Specifically, stellar evolution is simulated with initial metallicities (absolute units) of $Z \in \{ 10^{-5}, 10^{-4},$ $0.001,$ $0.002,$ $0.003,$ $0.004,$ $0.006,$ $0.008,$ $0.01,$ $0.014,$ $0.02, 0.03, 0.04\}$.} and implements the mass loss prescription from \citet{dejager_1988} with a transition to \citet{vink_2001} for OB stars. Furthermore, if the hydrogen surface abundance is below 0.4 and the surface temperature above $10^4$ K, \textsc{bpass} transitions to \citet{nugis_2000}. The stellar wind mass loss prescriptions are scaled to other metallicities with $Z_\odot=0.020$ and $\dot{M}(Z) = (\dot{Z_\odot})(Z/Z_\odot)^\alpha$, where $\dot{M}$ is the stellar mass loss rate and $\alpha = 0.5$, except for OB stars when $\alpha=0.69$. While the scaling of solar metallicity will matter, we perform the analysis in this work using absolute metallicity values.
We discuss the impact of stellar wind mass loss on the PISN rate in Section \ref{ssec:stellar_modelling}.

The use of a stellar evolution code gives access to the internal structure of the star at the moment of collapse, allowing us to explore the implementation of multiple PISN prescriptions. Over the last few years, several PISN prescriptions have been proposed. These either use the helium (He) core or the Carbon-Oxygen (CO) core of the progenitor to determine if the system undergoes a PISN.
The standard \textsc{bpass} code uses the prescription by \citet{woosley_2007}, which is based on the final helium core with masses between $64 \msun$ (inclusive) and $133 \msun$ (exclusive). Because binary interactions can strip the star in the later stages of its evolution and change its explodability \citep{schneider_2021, Schneider+24}, we also implement the prescription of \citet{marchant_2019}, where stars with $M_\text{CO} \geq 60$ experience disruptive pair-instability. Since no upper limit is given, we use the upper helium core mass limit from \citet{woosley_2007}, which results in the total fiducial PISN metallicity bias function. We present this metallicity bias function in Appendix \ref{app:otherPISN_prescriptions}, comparing it to other choices of PISN formation models.
Although this model provides a similar metallicity bias function to \citet{Briel+21}, this rate is overall a factor 2 higher due to the removal of the requirement for an oxygen-neon core to be present at the moment of explosion. Appendix \ref{app:otherPISN_prescriptions} shows that using the He or CO core only minimally influences the selected models.

In the \textsc{bpass} models, PISNe can result from either the explosions of the primary or secondary stars in a binary, from a product of their mergers, or from the QHE channel. We discuss the rates of each possible formation pathway for PISNe as a function of redshift in Appendix \ref{app:formation_pathways}.

Furthermore, we also use the detailed stellar structure to link the PISN progenitor models to the theoretical PISN light curves from \citet{Kasen+11} based on their final envelope and core masses in their Table 1. They provide 3 classifications of progenitors models: blue supergiants, red supergiants (RSGs), and helium stars.
Since the BPASS metallicities only reach down to $10^{-5}$, we do not find any matches to their zero-metallicity blue supergiant models. The light curves should be minimally impacted by the metallicity of the progenitor model \citet{Kozyreva+14} and the effect of metallicity is included in the stellar evolution of the progenitor models.
Moreover, most models are stripped from all hydrogen when reaching the PISN regime and match a helium star model. We discuss further how each PISN progenitor modelled in \textsc{bpass} is matched to a lightcurve and show the rates of PISN that would produce each lightcurve in Appendix \ref{app:light_curves}.

The models matching the red supergiant progenitors only match below $Z=0.007 Z_\odot$ or because the binary merges close to the moment of explosion, making it impossible for the merger remnant to lose its accreted hydrogen envelope in time. In Section \ref{ssec:results-by-pathway}, we discuss the formation pathways for PISN produced by \textsc{bpass} in more detail.

\subsection{Cosmological simulation}

The IllustrisTNG project is a suite of large-volume cosmological magnetohydrodynamical simulations, containing sub-grid models for many important astrophysical processes including supermassive BH formation, growth, and feedback, stellar evolution, stellar feedback, and enrichment of the interstellar medium \citep{TNG1, TNG2, TNG3, TNG4, TNG5}. In this study, we use the TNG50-1 simulation as it has the highest resolution ($8.5 \times 10^4 \msun$ per baryonic particle), allowing it to resolve internal metallicity distributions of galaxies to fine detail, and capture small, metal-poor galaxies that may be likely sites for PISN progenitors.

This simulation is based on the moving-mesh code \textsc{arepo} \citep{Springel10}, in which mesh-generating points flow freely under the effects of gravity and are used to construct a Voronoi tessellation at each timestep. This methodology combines aspects of both Eulerian mesh-based methods, which tend to overmix metals via advection between cells \citep{Sarmento+17}, with Lagrangian particle-based methods in which mixing is under-resolved \citep{Wiersma+09}. Understanding whether the interstellar medium of TNG galaxies is realistically well-mixed is an ongoing subject of research \citep[e.g.][]{metha_2023}. 

Sixteen different snapshots were downloaded from this simulation, ranging from $z=0$ to $z=10$.\footnote{Snapshots 4, 6, 8, 11, 13, 17, 21, 23, 25, 29, 33, 40, 50, 67, and 99 -- all available for public download at \url{https://www.tng-project.org/data/downloads/TNG50-1/}.} We follow \citet{Metha+20} and define all \emph{galaxies} in each snapshot to be collections of particles contained in dark matter subhalos that would be visually inseparable. This definition serves two purposes: it prevents the issue of dark matter overdensities embedded in larger subhalos from being classified as independent systems \citep{TNG2}, and is more harmonious with the definition used by observational astronomers. We define two subhalos to be visually inseparable if their half-star radii overlap -- see Section 2.2 of \citet{Metha+20} for further details.

\begin{figure}
    \centering
    \includegraphics[width=0.49\textwidth]{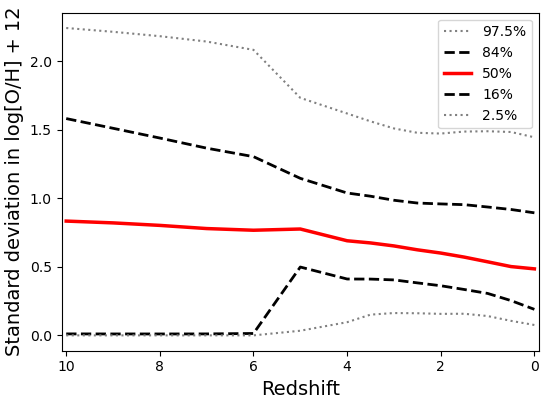}
    \caption{The standard deviation in the gas-phase metallicities of star-forming galaxies in the TNG50 simulation as a function of redshift. Only well-resolved galaxies, defined as galaxies containing more than 100 gas particles, are considered in this plot. The solid red line denotes the median standard deviation of star-forming TNG galaxies. Dashed black lines represent the 16$^\text{th}$ and 84$^\text{th}$ percentiles of the standard deviation of the metallicity at all redshifts. Dotted grey lines enclose $95$ per cent of galaxies. From this plot, it can be seen that galaxies at all redshifts show a wide variety of internal metallicity profiles. Furthermore, typical star-forming galaxies do not have uniform metallicities throughout -- instead, the median standard deviation of the gas-phase metallicity of galaxies at $z=0$ is $0.5$ dex, and this value increases with redshift.}
    \label{fig:TNG_standard_deviation}
\end{figure}

Figure \ref{fig:TNG_standard_deviation} shows the standard deviation of the metallicity present within star-forming galaxies in the IllustrisTNG simulation as a function of cosmic time. The red solid line shows the median variability of metallicity (in units of dex) for galaxies in TNG50-1, while the dotted and dashed curves show the standard deviations for 1$\sigma$ and 2$\sigma$ outliers. Together, these curves illustrate how the internal variability of metallicity in the population of star-forming galaxies changes with cosmic time.

From this graph, we see that star-forming galaxies are not predicted to have the same metallicity everywhere. At a redshift of zero, the median standard deviation of metallicity in star-forming galaxies is $\sigma = 0.5$ dex. As redshift increases, this value increases also, implying that higher-redshift galaxies are more chemically inhomogeneous, on average, than local ones.

Next, for each galaxy in each snapshot, we compute the rate of PISN formation in two different ways. In both cases, we neglect any delay times between star formation and PISN production and assume that the rate of PISN is proportional to the star formation rate. Such an assumption is reasonable as the time delays for PISN production are predicted to be 5 Myr at most \citep{Briel+21}, which is shorter than the lifetime of a star-forming HII region \citep{tremblin_2014, xiao_2018}.

In the first case, we assume that galaxies have uniform metallicities throughout, and compute the rate of PISN ($R_{\text{PISN}}$) using the following equation:

\begin{equation} \label{eq:rate}
    R_{\text{PISN}} = \kappa(Z) \times SFR,
\end{equation}

where $\kappa(Z)$ is the metallicity bias function of choice (as shown in Figure \ref{fig:PISN_bias_functions}).

In the second case, we account for chemical inhomogeneities within gas cells by computing individual PISN rates for each gas cell before summing them up to get the total PISN rate for each galaxy:

\begin{equation} \label{eq:cell_wise_rate}
     R_{\text{PISN}} = \sum_{\text{all gas cells}} \kappa(Z_{\text{cell}})\times SFR_{\text{cell}}.
\end{equation}

\subsection{Observability of PISNe}

To compute the observable PISN fraction, we first need to translate our rates from the rest-frame comoving volumetric units that are used internally in the IllustrisTNG simulation code($R_{\text{PISN}}$, in units of cMpc$^{-3}$ yr$^{-1}$), to rates in square degrees per year that are more useful for observers, accounting for the effects of time dilation ($\Sigma_{\text{PISN}}$, in units of deg$^{-2}$ yr$^{-1}$).

We assume a flat $\Lambda$CDM cosmology, adopting values of $\Omega_m=0.31$, $\Omega_\Lambda=0.69$, and $H_0=67.77$ km/s/Mpc \citep{Planck15}. Under this model, the size of a comoving volume element d$V_C$ within a section of solid angle d$\Omega$ and a redshift interval d$z$ is:

\begin{equation*} \label{eq:comoving_volume_element}
    \text{d}V_C = \frac{c \  {D_A(z)}^2 \ (1+z)^2}{H_0 \sqrt{\Omega_m(1+z)^3 + \Omega_\Lambda}} \text{d}\Omega \text{d}z,
\end{equation*} 
where $D_A(z)$ is the angular diameter distance at redshift $z$. Using this, we can compute the observed rate per solid angle between two redshift $z_1$ and $z_2$ from the predicted rate per comoving volume using the following equation:

\begin{equation*} \label{eq:convert_to_per_deg2}
    \Sigma_{\text{PISN}} = \left( \frac{\pi}{180} \right)^2 \int_{z_1}^{z_2} \frac{c \  {D_A(z)}^2 \ (1+z)}{H_0 \sqrt{\Omega_m(1+z)^3 + \Omega_\Lambda}} R_{\text{PISN}} \text{d}z,
\end{equation*}
noting that one factor of $1+z$ has been absorbed to account for time dilation.

Once the rate of PISN in units of yr$^{-1}$deg$^{-2}$ is known, we can then compute the number of PISN expected by an observational campaign ($N_{\text{obs}}$). To do this, we follow the methodology of \citet{Moriya+21}:
\begin{equation}
    \label{eq:nobs}
    N_{\text{obs}} = \epsilon \Sigma_{\text{PISN}} \tau_s A_s
\end{equation}
where $\tau_s$ is the survey's duration, $A_s$ is the area of the survey, and $\epsilon$ is the total fraction of PISNe that are visible to the telescope, hereafter referred to as the \textit{discovery fraction}. For a given survey using a specified telescope, the discovery fraction of PISNe can be estimated by performing a mock transient survey on PISNe produced with a specified lightcurve, at a range of redshifts. We provide more details on how this is done in Section \ref{sec:visible_rates}.

\section{Model predictions} \label{sec:results}

\subsection{Rates of PISN through cosmic time}
\label{ssec:results-rates}

\begin{figure}
    \centering
    \includegraphics[width=0.45\textwidth]{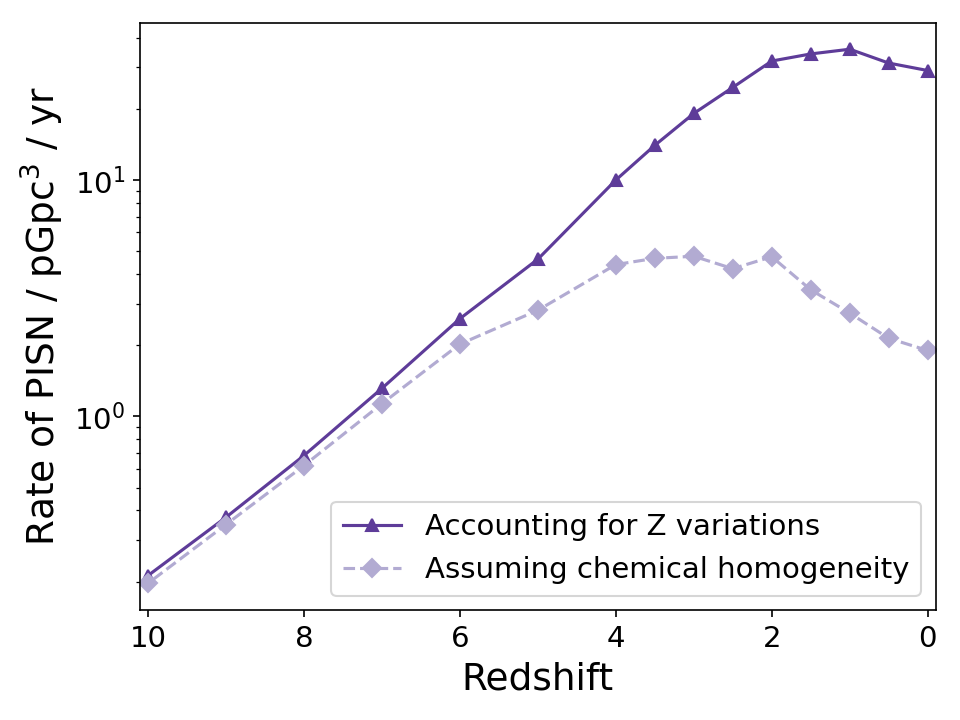}
    \caption{Rates of PISN as a function of comoving volume, not accounting for time dilation. At high redshifts ($z \gtrsim 6$), the two functions seem to trace each other well. However, at lower redshifts, the BPASS model that accounts for metallicity inhomogeneities within galaxies produces PISN much more efficiently, peaking at a redshift of $z=1-2$. At low redshift ($z=0-1$), the model that accounts for metallicity inhomogeneities is an order of magnitude more efficient at producing PISN than the other model.}
    \label{fig:volumetric_rate}
\end{figure}

Adding together contributions from all galaxies, we plot the expected number of PISN in each snapshot of the simulation in units of Gpc$^{-3}$ yr$^{-1}$ in Figure \ref{fig:volumetric_rate}. At all redshifts, accounting for chemical inhomogeneities in galaxies increases the expected PISN rate. At high redshifts ($z \gtrsim 6$), the rate of PISN computed using the assumption that metallicities are uniform is very similar to the rate computed accounting for internal metallicity variations. This is in the middle of the Epoch of Reionisation, where the universe is still mostly composed of pure atomic hydrogen. Star-forming galaxies in this period tend to have an average metallicity of $2 \times 10^{-3}$, making it possible for them to be PISN hosts without the presence of low-metallicity pockets of dense, star-forming gas. In this era, the spread of metallicities inside galaxies is still large (Figure \ref{fig:TNG_standard_deviation}) but is limited to low metallicities. 

Below $z=6$, the behaviour of the rates as a function of cosmic time changes for the two models. Under the assumption that galaxies have uniform metallicities throughout, the rate of PISN decreases, as the universe becomes more enriched, and it becomes less likely to find galaxies with metallicities below $\sim 0.1 \Zsun$. However, when the presence of metallicity variations within galaxies is accounted for, we find that the volumetric rate continues to rise to a redshift of $z\approx1.5$. This behaviour can be explained as a consequence of two competing effects: the enrichment of the Universe lowers the proportion of stars that have sufficiently low metallicity to become PISN, but the increased star formation rate at these epochs acts to oppose this effect. This theory predicts that low metallicity PISN may occur in galaxies with average metallicities higher than the theorised PISN cutoff. We will examine the population statistics of these host galaxies in more depth in Section \ref{sec:host-gals}.

The non-uniform PISN rate keeps increasing until $z \approx 1$ to a peak rate of 35 PISN Gpc$^{-3}$ yr$^{-1}$ due to the metallicity spread inside galaxies. However, eventually, the spread in metallicity inside a galaxy becomes smaller, as shown in Figure \ref{fig:TNG_standard_deviation}, and the mean metallicity of galaxies becomes near solar. These two effects contribute to a decrease towards $z=0$ and a rate of 29 PISN Gpc$^{-3}$ yr$^{-1}$. This rate is an order of magnitude higher rates than the uniform-metallicity with 1.9 PISN Gpc$^{-3}$ yr$^{-1}$ at $z=0$, and a peak rate of 4.7 PISN Gpc$^{-3}$ yr$^{-1}$ around $z\sim2$, at a higher redshift than the non-uniform distribution.

\subsection{PISN formation pathways}
\label{ssec:results-by-pathway}

\begin{figure}
    \centering
    \includegraphics[width=0.45\textwidth]{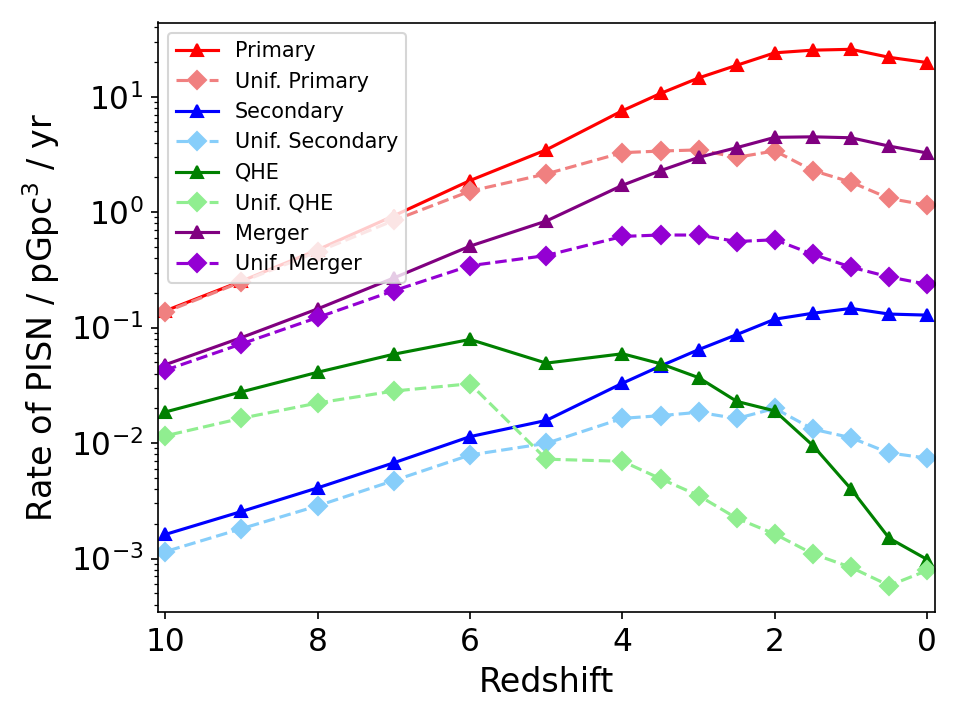}
    \caption{The cosmic PISN rate per progenitor model, where primary and secondary indicate the initially more massive or less massive star, respectively, exploding as a PISN. In a merger model, the initially more massive star merges with a main-sequence companion. Solid lines and triangles indicate models where chemical inhomogeneities in PISN host galaxies are considered. Dashed lines with diamonds indicate models where PISN host galaxies are assumed to have uniform metallicities throughout.}
    \label{fig:formation_pathway_rates}
\end{figure}

Since most massive stars have a companion, binary interaction can affect their final stellar structure and the ejecta material in the supernova \citep{laplace_2020, schneider_2021}. While \textsc{bpass} does not consider the effects of rotation in the stellar models, it does consider the effect of mass transfer and stellar mergers. We split all PISN progenitors models in the fiducial model selection by their formation channel in Figure \ref{fig:formation_pathway_rates}.

In the \textsc{bpass} models, there is no single star contribution to the PISN rate due to the multiplicity fraction above $65 \msun$ being 1 or higher. As such, all stars above this mass are in binaries.
The primary channel dominates the rate due to their per-definition higher zero-age main-sequence (ZAMS) masses than the secondary channel. For both the primary and secondary channels, the progenitor stars have ZAMS masses around 100-300 $\msun$.
Similarly, very few quasi-homogeneous stars result in PISN in \textsc{bpass} due to them starting as the initially less massive star and having to accrete tens of $\msun$ to become a chemically homogeneous star.

A more unique formation channel of PISN progenitors is the merger model, which goes to higher metallicities than any other channel, forming about 10 per cent of all PISNe at $z=0$. This formation channel has been suggested as a formation pathway for hydrogen-rich PISN, and it has been suggested that this formation pathway may dominate the local PISN rate \citep{vigna-gomez_2019}. Although in \textsc{bpass} a merger can occur at any evolutionary phase of the primary star, the companion will always be a hydrogen-rich main sequence star per definition. This adds fresh hydrogen to the outer layers of the star. However, this is quickly stripped away due to increased stellar winds in the post-merger star. In total, only 1 per cent of PISN are tagged as hydrogen-rich progenitors, which we discuss in more detail in Section \ref{sec:discussion}. Since the stellar winds for very massive stars implemented in \textsc{bpass} are weaker than current literature suggests \citep[][and see Section \ref{sec:discussion}]{Vink22}, more material would potentially be removed, further reducing the chances of a hydrogen-rich PISN.

\begin{figure}
    \centering\includegraphics[width=0.45\textwidth]{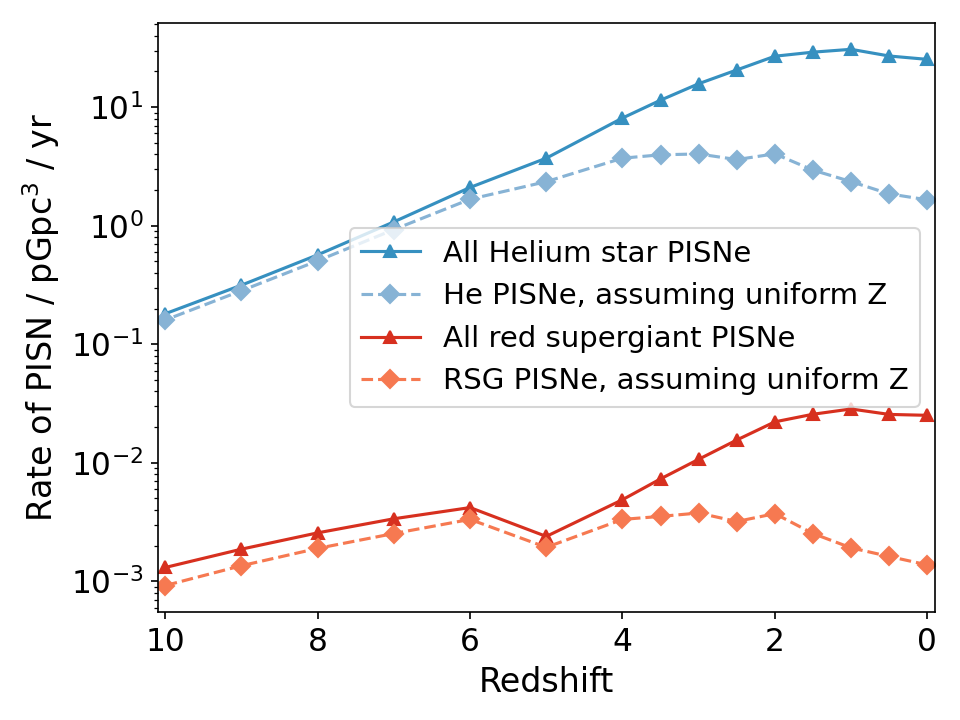}
    \caption{The cosmic PISN rate split by helium (blue) or red supergiant (red) light curve from \citet{Kasen+11}. We show the detailed galaxy metallicity distribution (solid) and uniform galaxy metallicity (dashed) over redshift.}
    \label{fig:rates_by_supertype}
\end{figure}

Shifting our focus to the progenitor models and to which light-curve the model is tagged, we show the rates of PISN from helium cores (He) and red supergiants (RSGs) in Figure \ref{fig:rates_by_supertype}. In both cases, the main results of Figure \ref{fig:volumetric_rate} are seen; namely, that the PISN rates are larger when chemical inhomogeneities are accounted for, increasing the PISN rate by over an order of magnitude at $z=0$. For the RSG channel, PISNe are limited to extremely metal-poor progenitors $Z \leq 10^{-5}$, and such PISNe are 2-3 orders of magnitude less likely to occur than the helium star PISN. We show more detailed rates of each light curve from the helium and RSG models in Appendix \ref{app:light_curves}.

The fraction of bright PISN remains mostly constant above $z=4$ because the metallicity bias function is similar at low metallicity. Only when the universe becomes sufficiently enriched do the relative contributions start to change and does the He70 model become dominant. These non-super-luminous PISN models could provide an additional probe for the pair-production regime and PISN mass gap, depending on whether they have been observed in existing surveys.

\subsection{Metallicity distribution of PISN progenitors}

The metallicity thresholds for PISNe in \textsc{bpass} lies at $Z \approx 0.43 Z_\odot$, which is slightly higher than the often used ${\sim} 0.3 Z_\odot$ for PISNe \citep[e.g.][]{langer_2007, farmer_2019, hendriks_2023}. As shown in Figure \ref{fig:formation_channels}, only merging models are contributing to the highest metallicity bin and have been suggested as a unique formation channel of PISNe progenitors at higher metallicities. The other formation channels are restricted to $Z \lesssim 0.29 Z_\odot$.

\begin{figure}
    \centering
    \includegraphics[width=0.45\textwidth]{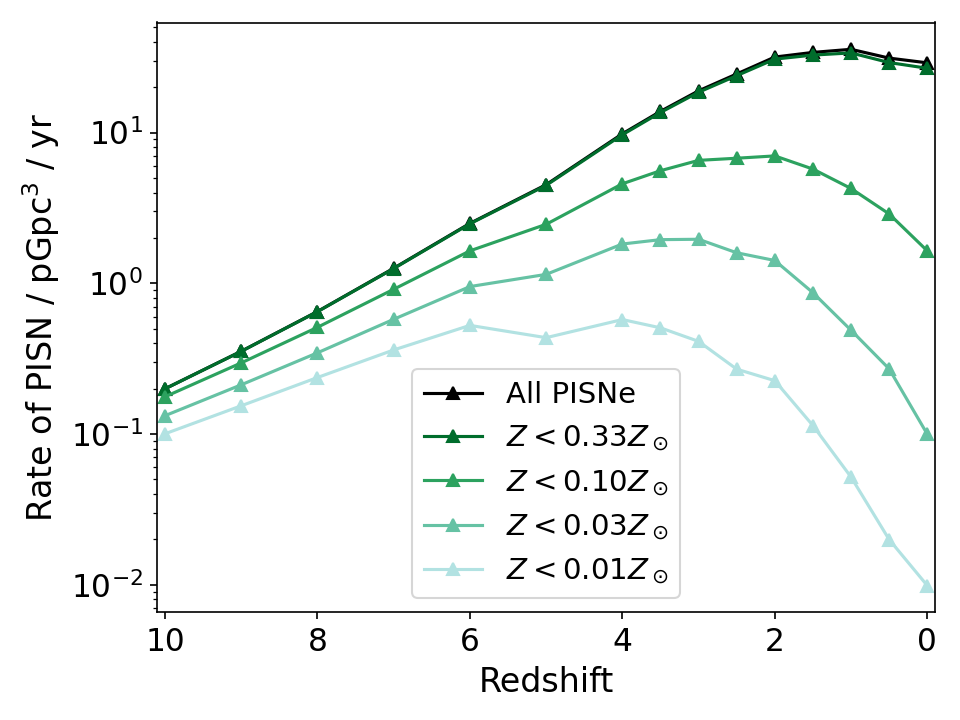}
    \caption{The rate of PISN with progenitor stars below metallicities of $1/3 Z_\odot$, $1/10 Z_\odot$, $1/30 Z_\odot$ and $1/100 Z_\odot$ throughout cosmic time. In this Figure, all results are computed accounting for chemical inhomogeneities inside host galaxies.}
    \label{fig:Z_thresholds}
\end{figure}

To investigate the sensitivity of our results to the \textsc{bpass} thresholds of galaxy metallicity, we investigated the metallicity distribution of PISN progenitor stars throughout cosmic time. Using our fiducial model, we plot the rate of PISN with metallicities below $1/3 Z_\odot$, $1/10 Z_\odot$, $1/30 Z_\odot$ and $1/100 Z_\odot$ in Figure \ref{fig:Z_thresholds}. While these thresholds are not motivated by any of the theoretical models investigated (Appendix \ref{app:otherPISN_prescriptions}), we include them to aid in comparison with other PISN models with lower metallicity thresholds. We find that almost all PISN produced in our modelling framework have metallicities lower than $1/3 Z_\odot$. PISNe with $Z < 0.1 Z_\odot$ peak at a redshift of $z=2$, compared to $z=1$ for the total PISN population. At $z=2$, PISNe with $Z < 0.1 Z_\odot$ make up $22\%$ of the PISN population. 
Using a stricter threshold of $1/30 Z_\odot$, the peak redshift of PISN formation shifts to $z=3$; and at this redshift, $10\%$ of PISNe have metallicities below this threshold. Finally, when PISNe with $Z < 0.01 Z_\odot$ are considered, the peak redshift is found to be $z=4$, where such PISNe make up $6\%$ of the population.

We note that the rate of PISNe predicted with metallicities below $1/10 Z_\odot$ when chemical inhomogeneities are accounted for is greater than the rate of PISNe predicted from our fiducial model when chemical inhomogeneities are ignored. This implies that accounting for chemical variations in the interstellar medium of galaxies is at least as important for predicting the rates of metal-poor transients as is having a precise metallicity threshold for their formation.

\subsection{Implications for transient surveys}
\label{sec:visible_rates}

Accounting for chemical inhomogeneities, the rate of PISN in the nearby universe is increased substantially -- by a factor of $>10$ for galaxies at $z \leq 1$, with a peak in the PISN rate shifted to lower redshift. Inspired by this result, we ask the question: \textit{how likely is it to see a PISN in the nearby Universe based on these rates?}

The visibility of each PISN produced in our simulation depends on its brightness, which in turn depends on the core mass and stellar evolution history of the star. \citet{Kasen+11} predict light curves for PISN with red supergiant, blue supergiant, and Wolf-Rayet progenitor stars, for which they compute rest-frame spectra of the resulting burst as a function of time. These can be convolved with a telescope filter and survey parameters to determine the observability of each light curve for those specific parameters.

PISNe are long-duration transients with light curves lasting hundreds of days, which require long-term observational surveys of the same field. Moreover, they occur at high redshift, with the predicted rate in this work peaking at $z\sim2$. Thus, deep field transient surveys are essential to discover them. We pick two observational surveys based on their deep fields, their long-term monitoring, and their current status to determine if we will or if we already have observed a potential PISN: the Hyper Suprime-Cam (HSC) transient survey \citep{Tanaka+12, Moriya+21}, and the Euclid Deep field survey (EDS) \citep{Moriya+22-Euclid}.

The HSC transient survey \citep{Moriya+21} observed 1.75 deg$^2$ consistently from late 2016 to early 2020, finding none to one super-luminous supernova lasting over 1 year. Using this data, they constrain the intrinsic PISN rate to less than 100 Gpc${}^{-3}$ yr${}^{-1}$ below $z=3$. The intrinsic rate for the non-uniform metallicity distribution at $z=0$ is already above this upper limit. However, the majority of local PISNe explode as He stars in our simulations, which is contrary to the equal distribution of light curve models assumed in \citet{Moriya+21}.

Figure \ref{fig:He_lightcurves} and \ref{fig:RSG_lightcurves} show the distribution of PISN rates corresponding to each light curve over redshift. Since PISNe from He progenitors are harder to detect than from RSG progenitors for the HSC survey, we calculate the number of PISNe this HSC transient survey would have seen based on the distribution of RSG and He progenitors. For each of the possible lightcurves of \citet{Kasen+11} (except for the He70 light curve\footnote{Data on the He70 lightcurve was not made publically available, and so rates of supernovae that would generate this lightcurve are excluded from this analysis. Such PISNe would not appear to be super-luminous, with a peak bolometric luminosity of $\sim 5 \times 10^{41}$ erg/s, and would likely not be detectable by HSC.}) at redshifts ranging from $0$ to $4$ in step sizes of $0.1$, we simulated a large number (thousands) of PISNe, in order to determine what fraction of PISNe of each lightcurve are visible to HSC as a function of redshift. 

Integrating over all lightcurves and all redshifts, using the rates presented in the previous Sections, we find that the total rate of PISNe that are visible to HSC is $0.0053$ PISNe per deg$^2$ per yr. If galaxies are instead assumed to be chemically homogeneous throughout, this number drops by more than an order of magnitude, with a predicted rate of $4.5 \times 10^{-4}$ deg$^{-2}$ yr$^{-1}$.

Because the footprint of the HSC transient survey is $1.75$ deg$^2$, and the duration of this survey was $3.25$ yr, we can use this value to compute the probability of finding a PISN detection in archival HSC survey data using Equation \ref{eq:nobs}. Under the assumption that galaxies are chemically homogeneous, the probability of finding a PISNe in this data is $0.0025$. Accounting for chemical inhomogeneities, this value rises to $0.030$ -- that is, there is only a $3$ per cent chance that a PISN was observed by HSC throughout its lifetime, which is consistent with a non-detection for this survey \citep{Moriya+21}.

The HSC survey only observed 1.75 deg$^2$, while the upcoming the EDS \citep{laureijs_2011, euclid_2022} will observe $53$ deg${}^{2}$ on a regular cadence for its six-year mission duration \citep[20 for the North field, 23 for the South field, and 10 for the Fornax field ][]{laureijs_2011, euclid_2022}. The near-infrared filters of EDS are better suited to observe PISNe from helium progenitors in numbers and to higher redshift \citep{Moriya+22-Euclid}. We use the observability fractions for each light curve from \citet{Moriya+22-Euclid} to determine the number of detectable PISN in the Euclid survey over redshift.

Using the same methodology as for HSC, we computed the total event rates for PISNe that would be observable by the EDS to be $0.023$ PISNe deg$^{-2}$ yr$^{-1}$ in the case where galaxies are assumed to have a uniform metallicity throughout, and $0.26$ PISNe deg$^{-2}$ yr$^{-1}$ for the case where chemical inhomogeneities in galaxies are accounted for. In both cases, the most likely redshift at which a PISN is expected to be detectable is $z\sim 1.8$, near the peak of star formation. Over a $53$ deg${}^2$ field of view and a six-year mission lifetime, the computed rates correspond to an expected number of PISN detections in EDS of $7.3$ for the chemically homogeneous case, rising to $83.3$ when the inhomogeneous nature of galaxies is considered -- higher by a factor of $11.4$. Such a large number of PISN detections would be useful to constrain PISN population statistics, giving valuable insights into the physical mechanisms that power such events.

\begin{figure}
    \centering
    \includegraphics[width=0.5\textwidth]{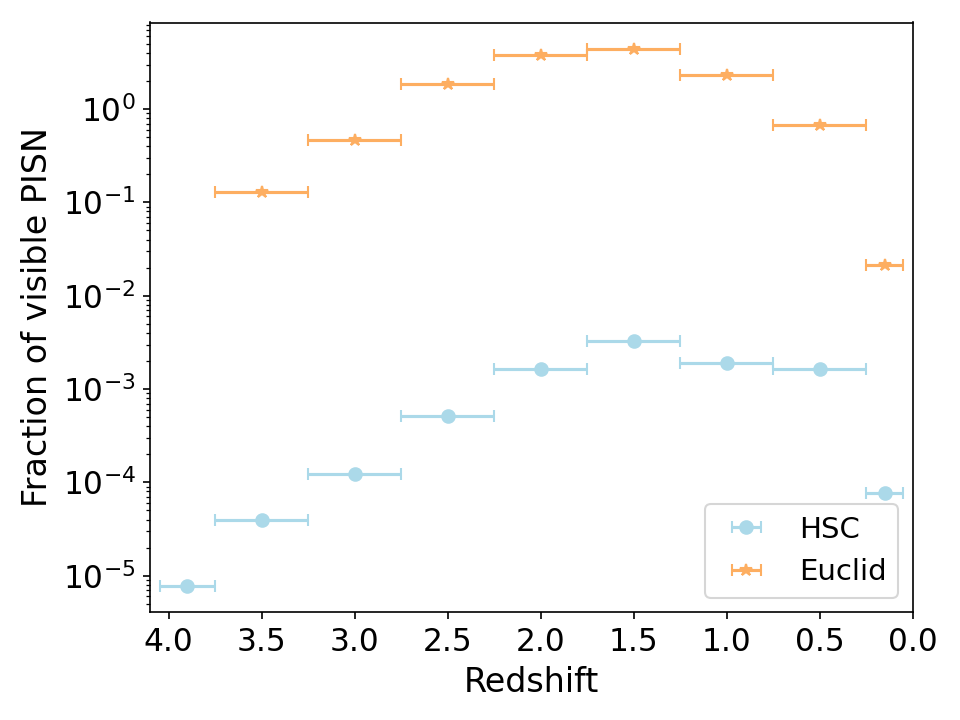}
    \caption{The number of PISNe that are detectable per year by the HSC transient survey (blue circles) or the Euclid Deep Survey (orange triangles) as a function of redshift. From this plot, we predict that most PISNe detectable by Euclid will occur around cosmic noon ($z\sim 2$).}
    \label{fig:vis_rates}
\end{figure}

In both cases, we find that most of the PISNe predicted by this work are not visible to either HSC or Euclid due to the majority of PISNe being too faint. At low redshifts, $(z=0.05-0.25)$, $14.3$ per cent of PISNe produced are visible to EDS, while only $1.4$ per cent of PISNe produced would be visible to HSC. By $z=4$, no PISNe are visible to either survey. Appendix \ref{app:detectable_fraction} shows the detection fraction of either survey as a function of redshift.
Combining the detection fraction with our estimations on the cosmic rate of PISNe, we can predict the detection rate of these telescopes for PISN discovery throughout cosmic time. We show this result in Figure \ref{fig:vis_rates}. We find that, for both surveys, the most likely redshift at which to discover PISNe is close to $z=2$, near the peak of star formation at cosmic noon. This Figure also highlights the level of improvement that EDS will have over HSC as a PISN detector.

\begin{figure}
    \centering
    \includegraphics[width=0.4\textwidth]{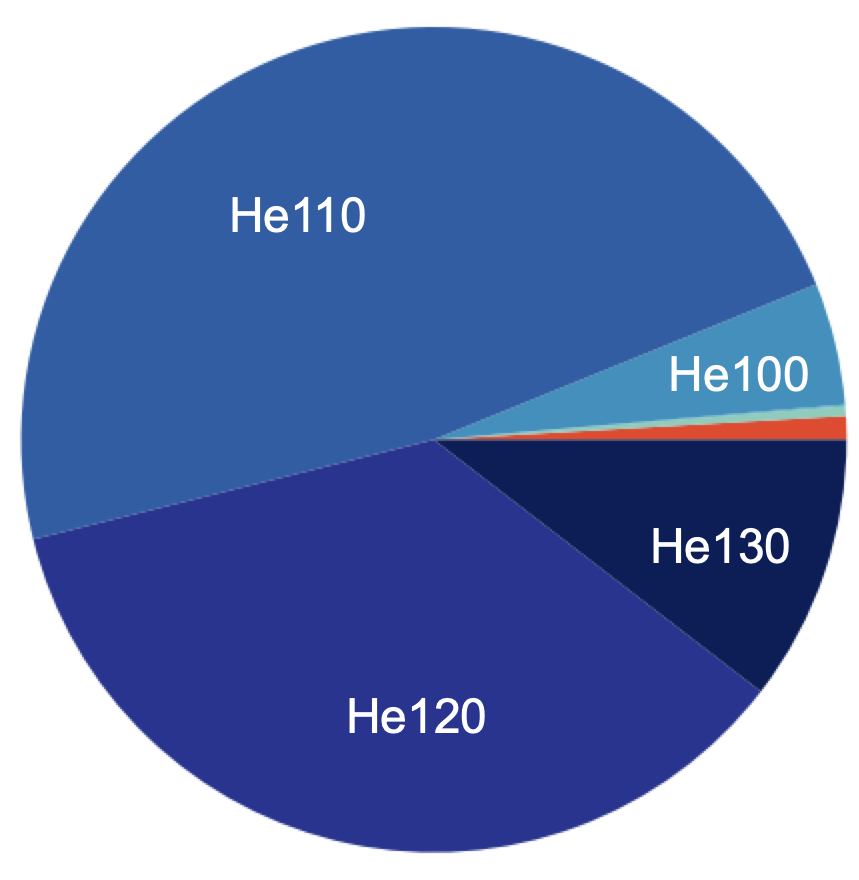}
    \caption{The fraction of PISN that are detectable by the Euclid Deep Survey, summing contributions from all redshifts explored. The vast majority of observable PISNe are expected to come from massive He stars (He110, He120 models of \citealt{Kasen+11}).}
    \label{fig:vis_pie}
\end{figure}

Using our modelling framework, we can also make predictions for which class of PISN the Euclid Deep survey is the most likely to see. We illustrate this breakdown in Figure \ref{fig:vis_pie}.
Considering contributions from all redshifts, and incorporating both the brightness data of each light curve and their predicted frequency, we discover that the most likely light curves to be visible to EDS will originate from massive He stars, with core masses of $100\msun$ ($4.8$ per cent of detectable PISNe), $110\msun$ ($47.7$ per cent), $120\msun$ ($35.7$ per cent), or $130\msun$ ($10.4$ per cent). All other bursts, including those originating from low-mass He-stars or red supergiants, make up less than $1.4$ per cent of the predicted detectable PISN sample. We, therefore, advise Euclid users to focus on hydrogen-free PISNe, which match the He110 and He120 light curves of \citet{Kasen+11} at redshifts of $z \sim 1.3-2.5$ for optimal chances of PISN discovery.

\section{Host galaxy properties}
\label{sec:host-gals}

PISN candidates have been observed in metal-poor star-forming, dwarf galaxies \citep{Schulze+23}, similar to the general super-luminous supernova population \citep{Schulze+21, Cleland+23}. In this Section, we make predictions about the population statistics of PISN host galaxies throughout cosmic time using our modelling framework.

\begin{figure}
    \centering
    \includegraphics[width=0.45\textwidth]{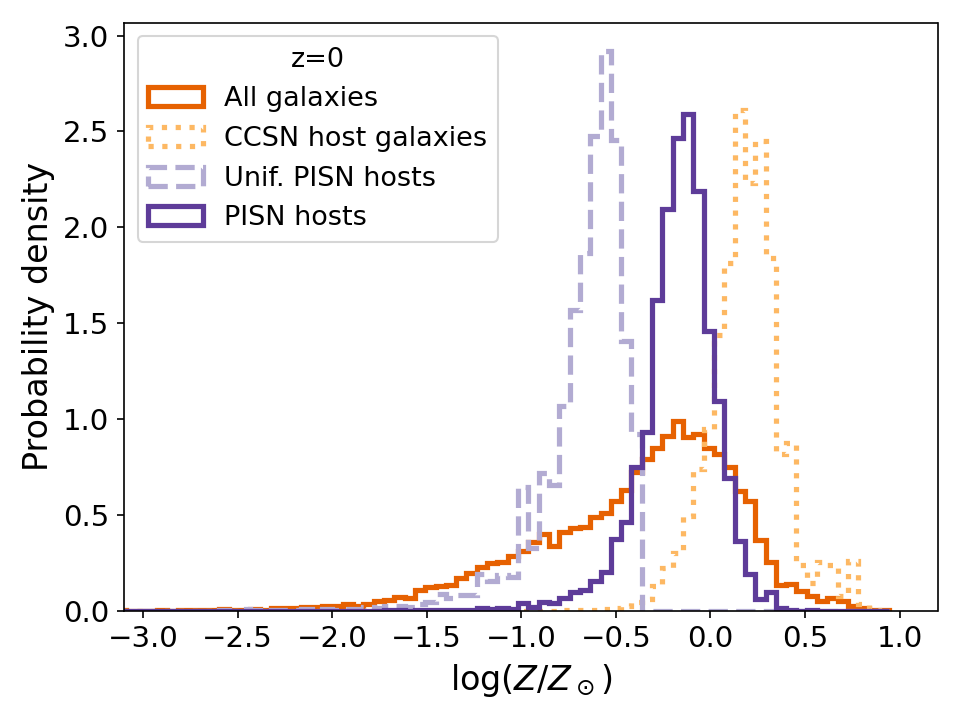}
    \caption{The metallicity distribution of PISN host galaxies at $z=0$, both accounting for metallicity inhomogeneities (solid dark purple line), and assuming galaxies have a constant metallicity throughout (dashed purple lines). For comparison, we also plot the metallicity distribution of all TNG galaxies, both unweighted (solid orange line), and weighted by the star formation rate of galaxies, which is a proxy for the CCSN rate (dotted orange line). When the inhomogeneous nature of local galaxies is considered, we find that the mean metallicity of PISN host galaxies is consistent with the mean metallicity of all galaxies at $z=0$, but with a much narrower spread.}
    \label{fig:z=0_host_metallicities}
\end{figure}

In Figure \ref{fig:z=0_host_metallicities}, we show the predicted metallicity distributions of four populations of host galaxies present in the IllustrisTNG simulation. For reference, we show the population statistics for all IllustrisTNG galaxies (solid orange line), and also for IllustrisTNG galaxies weighted by their core-collapse supernova (CCSN) rate (dotted orange line), which is roughly proportional to the star-formation rate weighted galaxy distribution. We see that galaxies that are actively forming new stars at $z=0$ tend to be more metal-rich than the average population of galaxies in the IllustrisTNG simulation. This can be understood to be a consequence of two well-studied galaxy scaling relations: (i) more massive galaxies have higher star-formation rates \citep[e.g.][]{Brinchmann+04}, and (ii) more massive galaxies have greater metallicities \citep[e.g.][]{Tremonti+04}.

On top of these distributions, we show the distribution of metallicities of TNG galaxies weighted by the rate at which PISNe are produced, (i) using the average metallicity of each galaxy and assuming that every galaxy has a uniform metallicity throughout (dashed light purple line), and (ii) by predicting the PISN formation rate in each gas cell associated with each galaxy, which captures chemical inhomogeneities within galaxies (solid dark purple line). By construction, when galaxies are assumed to have constant metallicities throughout, PISN host galaxies are, on average, $0.3$ dex more metal-poor than the average galaxy population, and $0.8$ dex more metal-poor the star-formation weighted average galaxy population. On the other hand, when internal variations of the metallicity inside galaxies are considered, the average metallicity of PISN host galaxies is $0.15$ dex higher than the average metallicity of IllustrisTNG galaxies at $z=0$, while remaining significantly ($0.35$ dex) less metal-rich than the CCSN host galaxy population. This can be understood to be a consequence of tension between two factors. More massive galaxies have higher star-formation rates and higher average metallicities. PISN host galaxies are biased to have higher star formation rates, but lower metallicities. Taken together, these two effects roughly cancel out, and the median metallicity of local PISN host galaxies (if they exist) is expected to be roughly aligned with the overall median metallicity of galaxies at $z=0$.


\begin{figure}
    \centering
    \includegraphics[width=0.45\textwidth]{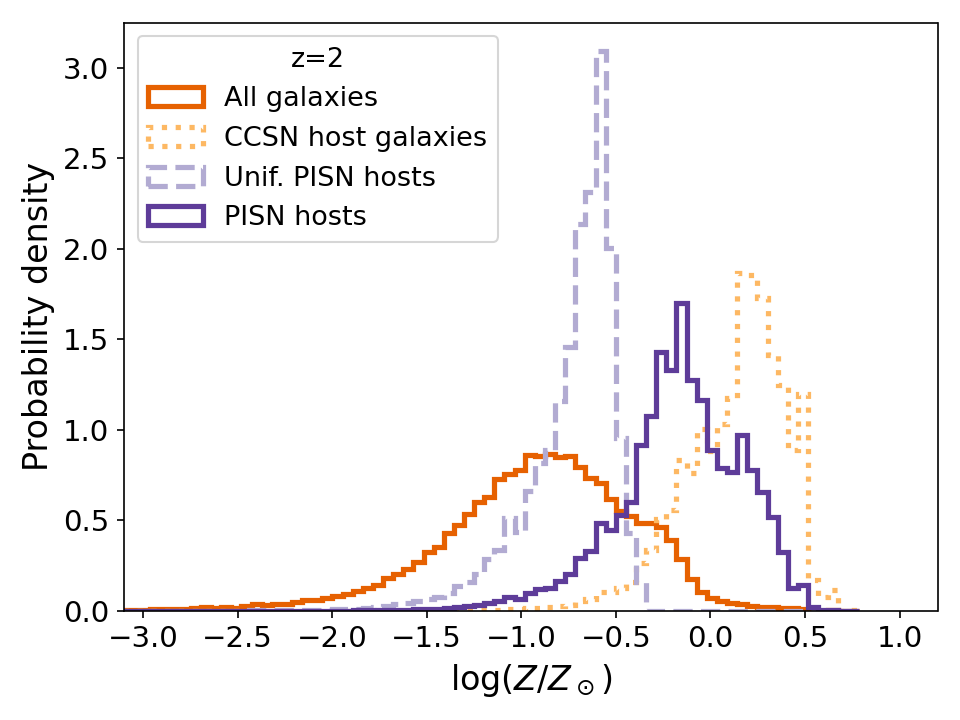}
    \caption{As in Figure \ref{fig:z=0_host_metallicities}, but at a redshift of $z=2$. When the inhomogeneous nature of IllustrisTNG galaxies is accounted for, we find that the host galaxies of PISNe have, on average, higher metallicities than most IllustrisTNG galaxies.}
    \label{fig:z=2_host_metallicities}
\end{figure}

In Figure \ref{fig:z=2_host_metallicities}, we show how this relationship is expected to change at cosmic noon, by examining the $z=2$ snapshot of IllustrisTNG.\footnote{Snapshot 33 of the TNG100-1 simulation.} At this redshift, we see that when the inhomogeneous chemical nature of TNG galaxies is accounted for, the median metallicity of PISN host galaxies is expected to be $0.75$ dex larger than the median galaxy metallicity, whilst still being $0.3$ dex lower than the CCSN host galaxy metallicity distribution. At this redshift, most metal-poor PISN progenitor systems are born in galaxies with average metallicities higher than the metallicity cutoff for PISN production.

\begin{figure}
    \centering
    \includegraphics[width=0.45\textwidth]{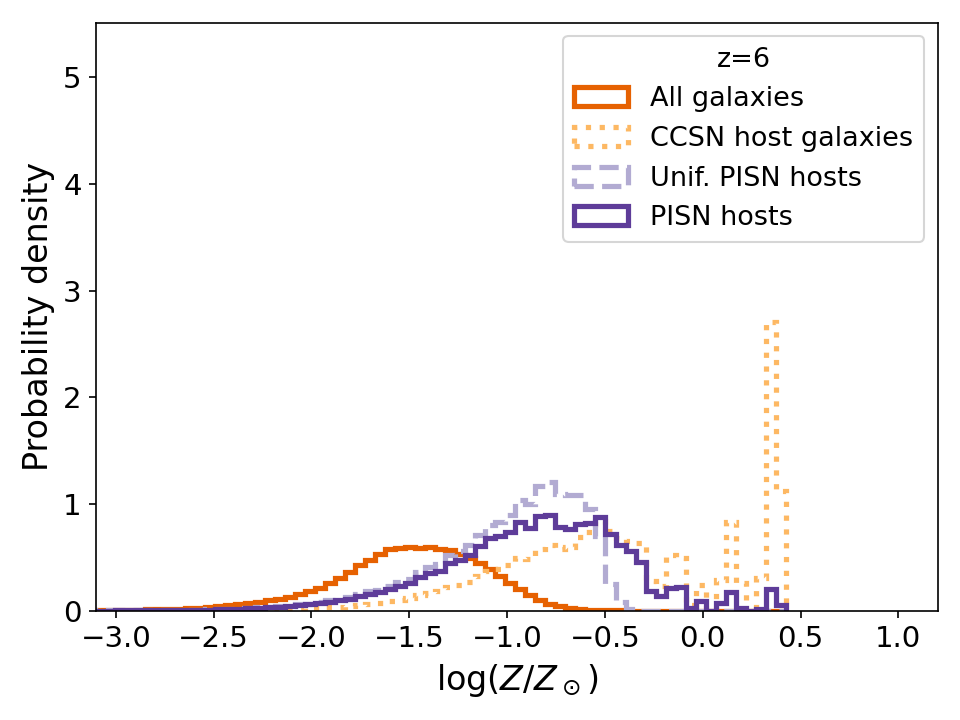}
    \caption{As in Figure \ref{fig:z=0_host_metallicities}, but at a redshift of $z=6$. In this epoch, accounting for chemical inhomogeneities in galaxies does not dramatically change the host galaxy metallicity distribution, in agreement with what was found in previous sections.}
    \label{fig:z=6_host_metallicities}
\end{figure}

Finally, we show the metallicity distribution of simulated PISN host galaxies compared to average (and CCSN host) galaxies at $z=6$, near the end of reionisation, in Figure \ref{fig:z=6_host_metallicities}. During this epoch, the average galaxy has a metallicity below the threshold for PISN production, but the average star-forming galaxy is still very metal-enriched. At this redshift, a large fraction ($38$ per cent) of galaxies are not metal-enriched, and so have a reported metallicity of $Z = 10^{-7.25} Z_\odot$, the metallicity floor used in the IllustrisTNG simulations. We find that only $4.4$ per cent of PISN form in completely unenriched galaxies, with the median metallicity of PISN hosts equal to $\log(Z/Z_\odot) = -0.87$ when metallicity fluctuations are accounted for. The median metallicity drops to $\log(Z/Z_\odot) = -0.96$ for uniform metallicity galaxies, while the fraction of PISN in completely unenriched galaxies increases slightly to $5.7$ per cent. In this epoch, the metallicity distribution of PISN host galaxies is similar when galaxies are considered to be chemically homogeneous when compared to the inhomogeneous case. This agrees with our findings in Section \ref{ssec:results-rates}, where we showed that at $z>6$, the rate of PISN is not increased by much when internal metallicity fluctuations in galaxies are accounted for. Indeed, we see that in the chemically inhomogeneous modelling case, only a few ($11$ per cent of) PISN host galaxies at this redshift have average metallicities higher than the cutoff metallicity for PISN formation.

At $z=6$, the PISN rate in both metallicity models more closely follows the metallicity distribution of the star formation rate weighted host galaxies with a preference for metal-poorer galaxies. The distribution of SFR and galaxy stellar mass at this redshift is wider and peaks at $5 \times 10^7 \msun$ instead of $\sim 10^9 \msun$ in the local universe. The peak SFR, on the other hand, remains around $0.1 \msun$ yr$^{-1}$, although is more widely spread around this value than at $z=0$. 

\begin{figure}
    \centering
    \includegraphics[width=0.45\textwidth]{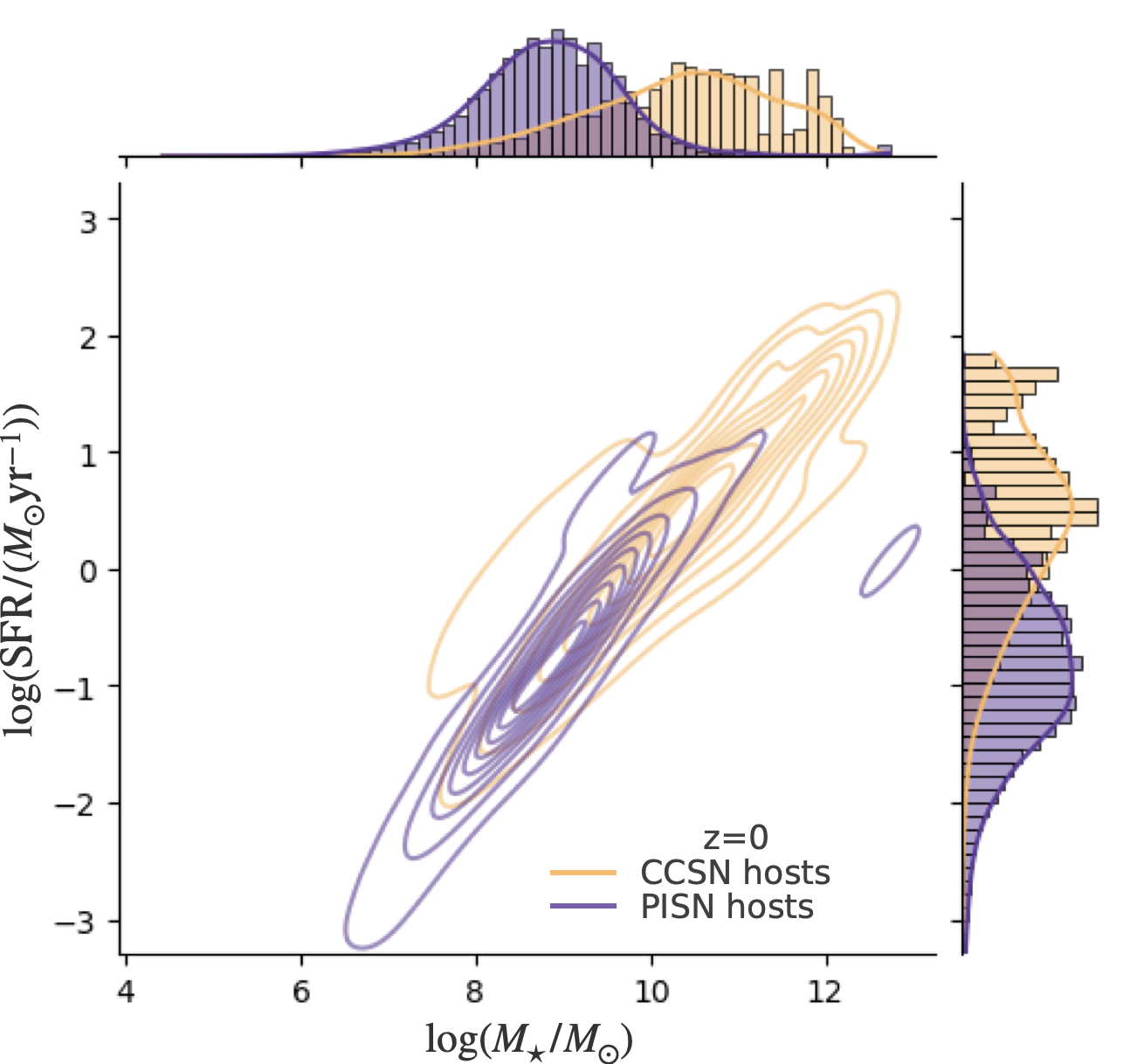}
    \caption{The distributions of host galaxy mass and star formation rate at $z=0$ in the IllustrisTNG simulation (orange) and the host galaxies weighted by their PISN rate (purple). The baryonic particle size is $8.5 \times 10^4 \msun$ for the TNG50-1 simulation used here. This limits the resolution/number of particles at the low mass regime. In this snapshot, PISNe masses and star formation rates are broadly consistent with the average IllustrisTNG galaxy properties.}
    \label{fig:host_SFR_M_z0}
\end{figure}

Next, we predict the mass and SFR distributions of PISN host galaxies throughout cosmic time. In Figure \ref{fig:host_SFR_M_z0}, we show the distributions of (i) the reference sample of CCSN host galaxies, and (ii) the population of PISN host galaxies at $z=0$ on the stellar mass-SFR plane. Consistent with Figure \ref{fig:z=0_host_metallicities}, we find that the predicted host galaxies of PISN at $z=0$ to be smaller and have smaller star-formation rates than typical CCSN host galaxies, with a median and 68$^{\text{th}}$ percentile range in stellar masses of $\log (M_*/M_\odot) = 8.86_{-0.80}^{+0.73}$, and star formation rates of SFR = $0.15_{-0.12}^{+0.72}M_\odot$yr$^{-1}$, compared to $\log (M_*/M_\odot) =10.4_{-1.22}^{+1.04}$ and SFR=$2.9_{-2.5}^{+13.0} M_\odot$yr$^{-1}$ for the reference population. In other words, at this redshift, the PISN host galaxy population lives in a subset of the general star-forming galaxy population with slightly lower masses and star formation rates than the average systems.

\begin{figure}
    \centering
    \includegraphics[width=0.45\textwidth]{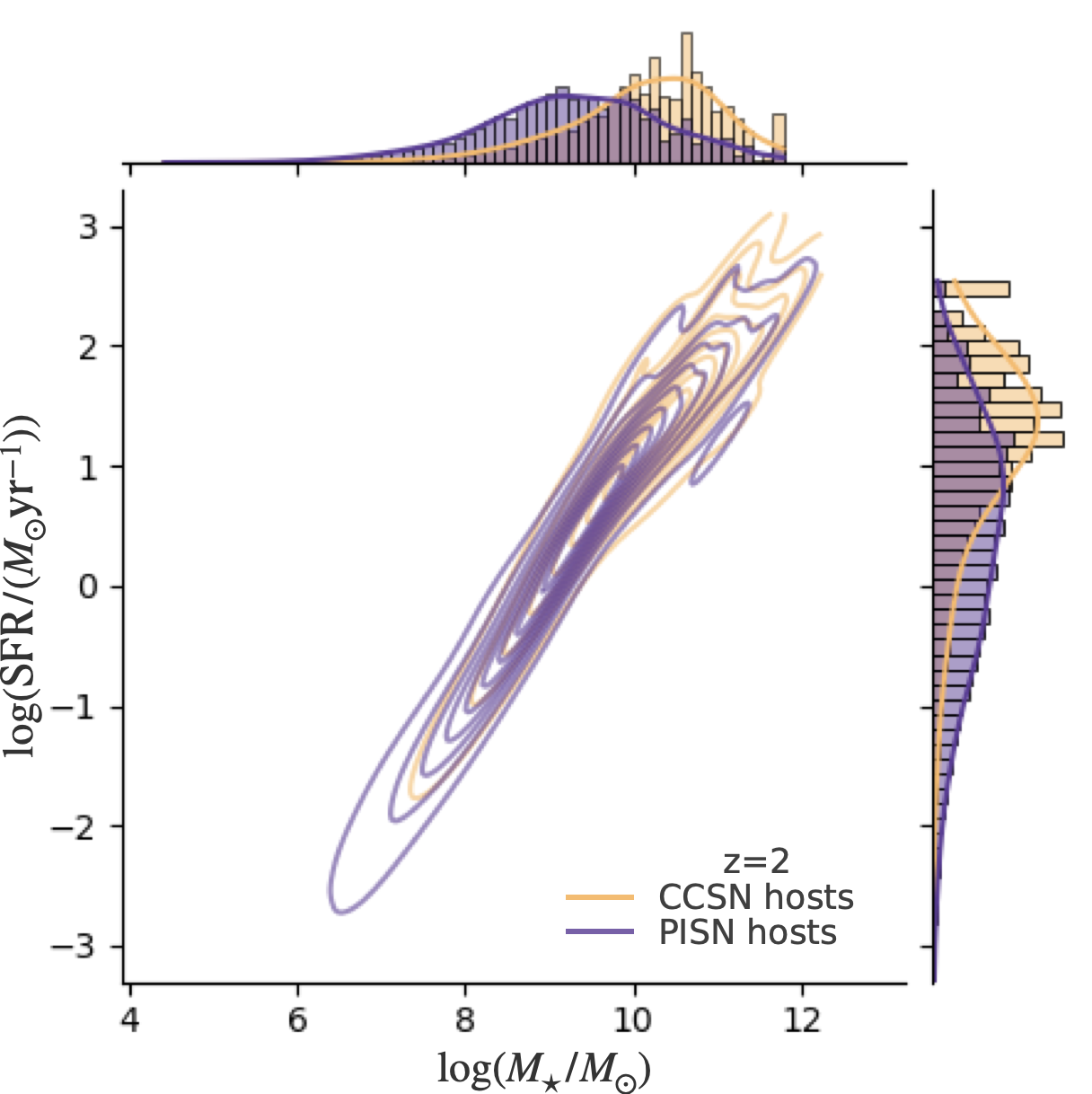}
    \caption{As in Figure 11, but with $z=2$. In this snapshot, PISNe tend to be hosted by more massive galaxies with higher star formation rates.}
    \label{fig:host_SFR_M_z2}
\end{figure}

At cosmic noon ($z=2$), we plot the distribution of PISN host galaxies on the $M_*$-SFR plane, compared to the reference sample of CCSN host galaxies, in Figure \ref{fig:host_SFR_M_z2}. In this epoch, $68$ per cent of PISN host galaxies have star formation rates of $2.82_{-2.65}^{+20.58}M_\odot$ yr$^{-1}$ and stellar masses of $\log (M_*/M_\odot) = 9.29_{-1.06}^{+1.01}$, compared to the significantly larger values of SFR = $18.1_{-15.8}^{+60.5}M_\odot$ yr$^{-1}$ and $\log (M_*/M_\odot) = 10.2_{-0.88}^{+0.77}$ computed for CCSN hosts. This shows that at cosmic noon, PISN host galaxies are expected to be systems capable of being seen by 8-meter class ground-based telescopes such as Keck.

\begin{figure}
    \centering
    \includegraphics[width=0.45\textwidth]{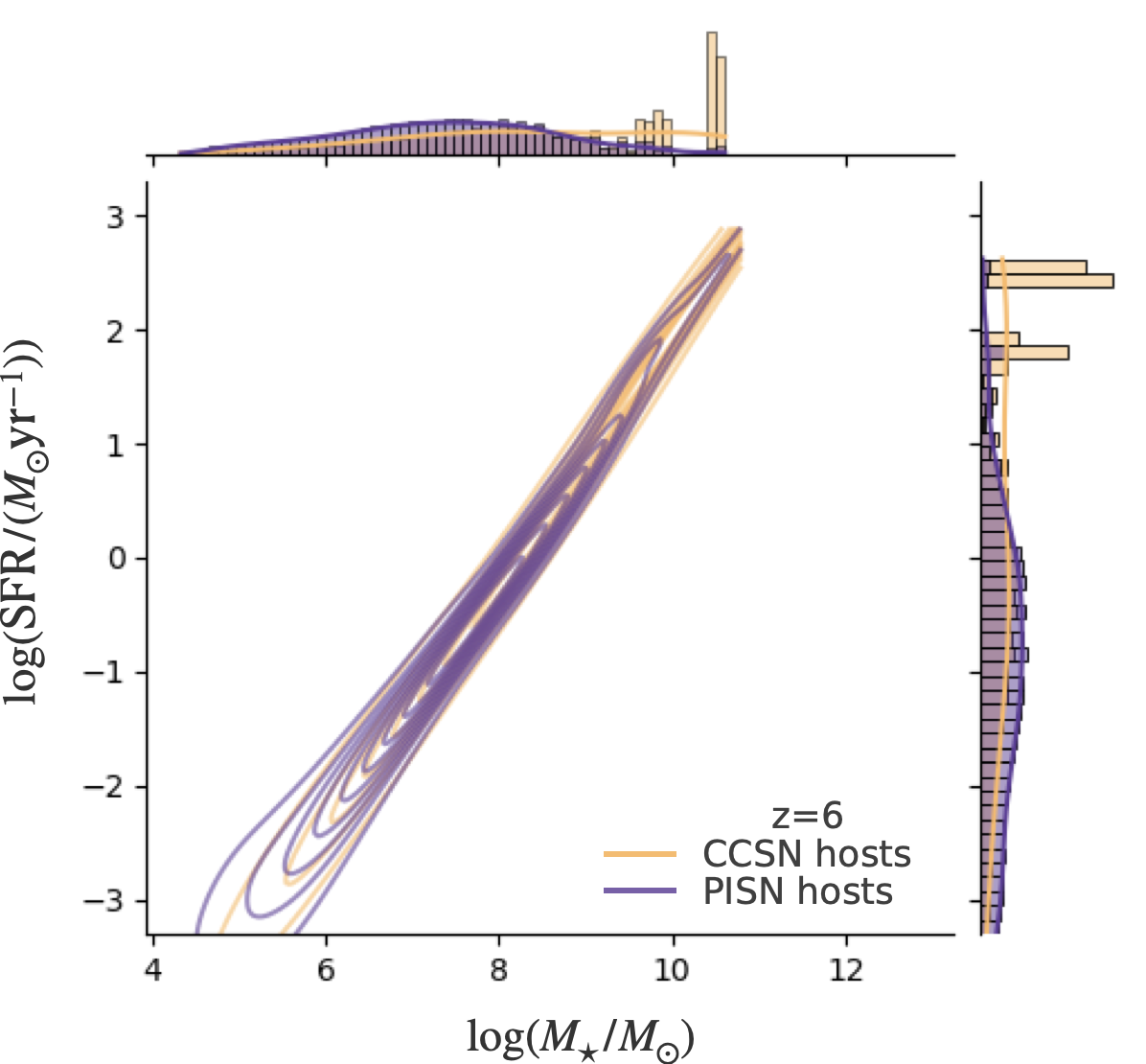}
    \caption{As in Figure 11, but with $z=6$. In this snapshot, PISN host galaxies are preferably found in starbursting systems, with star formation rates many orders of magnitude higher than the average IllustrisTNG galaxies.}
    \label{fig:host_SFR_M_z6}
\end{figure}

We show the predicted mass and SFR distributions of PISN host galaxies at $z=6$ in Figure \ref{fig:host_SFR_M_z6}. At this redshift, PISN host galaxies have star formation rates of approximately $2.09 \pm 0.1 M_\odot$ yr$^{-1}$ and stellar masses of $\log (M_*/M_\odot) = 7.28_{-1.60}^{+1.28}$, virtually indistinguishable from the overall population of CCSN galaxies at these redshifts. The limited enrichment of the universe at $z=6$ causes them to overlap and pushes the PISN and CCSN host galaxy populations apart towards lower redshifts, as seen in Figure \ref{fig:host_SFR_M_z0}. We note that at $z=6$ galaxies with such properties are outliers from the general population of galaxies, where star formation rates and stellar masses are typically much lower. Based on this prediction, we conclude that if a PISN was detected at a redshift of $z=6$ and its position was localised, its host galaxy would likely be able to be detected only using dedicated JWST follow-up. 

\section{Discussion} \label{sec:discussion}

While this is not the first work to make PISN rate predictions, this is the first PISN rate prediction using non-uniform metallicity internal galaxy distribution and detailed binary stellar evolution models. We compare our predicted rates against other predictions and observational limits, especially focusing on the limitations of observational boundaries.
We did not include Population III star formation, although this only provides a contribution above $z=6$, and therefore is not expected to significantly affect our local or observable PISN rate \citep{pan_2012, Tanikawa+23}.

\subsection{Stellar modelling}
\label{ssec:stellar_modelling}
The use of detailed stellar models comes with its own caveats and uncertainties, such as the implementation of stellar winds. In this work, we have not explored the impact of different stellar physics on the PISN, because the focus is on the impact of accounting for internal metallicity variations on small scales within galaxies.
However, stellar winds, nuclear rate, and rotation can drastically impact the location of the PISN regime by altering the size of the caron-oxygen core \citep[for a comprehensive review, see][]{Woosley+Heger21}.
\textsc{bpass} implements the stellar winds from \citet{Vink+01}, \citet{dejager_1988}, and \citet{nugis_2000} for their very massive stars, but does not include more recent mass loss prescriptions for very massive stars \citep{vink_2011} and red-supergiant mass loss \citep[see references in][]{Vink22}. In \citet{belczynski_2016}, the PISN metallicity limit is at $0.1 \Zsun$ with the implementation of the very massive star winds from \citet{vink_2011}. Above 80 to $90 \msun$, the observed wind mass loss rate is enhanced compared to the currently implemented prescription \citep{bestenlehner_2014} and might be metallicity independent \citep{smith_2023}. This reduces the possibility for very massive stars at higher metallicities to retain sufficient mass to reach the pair-production regime and can reduce the PISN regime to $Z_\odot/20$ for isolated single stars \citep{sabhahit_2023}. Depending on the metallicity dependence of these stellar winds, this also shifts more massive PISN progenitors to lower masses in the pair-production regime.

\citet{Tanikawa+23} implement a different set of stellar wind prescriptions, but only find a PISN rate that is 2 or 3 times smaller than the empirical model of \citet{Briel+21}.
The metallicity distribution in the star formation rate has a more significant impact, as shown in figure 10 in \citet{Briel+21}. At this time, the precise impact of stellar winds on the PISN rate is undetermined and is beyond the scope of this study.

Besides a decrease in the rate from stellar winds, PISN might also be avoided due to a decrease in the ${}^{12}\mathrm{C}(\alpha,\gamma){}^{16}\mathrm{O}$ nuclear reaction rate, which results in a larger ${}^{12}$C fraction. This shifts the PISN regime to higher masses \citep{Takahashi18, farmer_2019, farmer_2020}. As a result, this can decrease the PISN rate by 2 orders of magnitude in the local universe \citep{Tanikawa+23}. We performed a similar PISN mass range shift as \citet{Tanikawa+23} from 70-130 $\msun$ to $90-180 \msun$. We find a small decrease in the total rate by a factor of 2 at the peak, which is significantly smaller than the order of magnitude computed by \citet{Tanikawa+23}. 
Even with the increase of PISN from the non-uniform metallicity, this decrease makes our non-uniform PISN rate closer to ${\sim}17$ PISN Gpc$^{-3}$ yr$^{-1}$; higher than the uniform metallicity distribution and still observable with the Euclid Deep Survey.

Rotation can have a similar effect as the nuclear rates on the location of the PISN regime by stabilising the carbon-oxygen core leading to extended He cores \citet{Chatzopoulos+12} or through chemically homogeneous evolution \citep{duBuisson+20, Marchant+Moriya20, Woosley+Heger21}. \textsc{bpass} implements accretion-induced quasi-chemically homogeneous evolution, but does not include a contribution from tides, which is not considered in our modelling. Their effect has been explored for the long gamma-ray bursts and super-luminous supernovae populations in \textsc{bpass} \citep{chrimes_2020, ghodla_2023}. A more detailed analysis of its effect on the PISN population in \textsc{bpass} is required, because this channel could result in a significant boost to the PISN rate from the quasi-chemically homogeneous channel \citep{duBuisson+20}.

Incorporating any of these second-order considerations would result in a shift of the PISN regime to higher masses, which would decrease the intrinsic PISN rate due to the comparative rarity of more massive stars. However, it also affects the properties of the PISN progenitors. With more very massive stars undergoing pair-instability, more luminous events will contribute to the PISN rate. This could increase the number of PISNe visible to instruments such as Euclid, even though the intrinsic PISN rate is lowered. It is only through properly considering the variety of light curves that can be produced by PISN that observed PISNe rates can be meaningfully used to inform theories of PISN formation.

The predicted rates presented in this work are dominated by the $80 \msun$ He progenitor model. Thus, these changes can have dramatic effects on the intrinsic PISN rate, but many of these events are unobservable with Euclid. For example, the $3\sigma$ adjusted ${}^{12}\mathrm{C}(\alpha,\gamma){}^{16}\mathrm{O}$ nuclear reaction rate by \citet{Tanikawa+23} leads to 2 orders of magnitude decrease in the intrinsic rate, but only 1 order of magnitude in the total number of detected PISN by Euclid. Although they do not provide relative fractions of the \citet{Kasen+11} light curves, this indicates a reduced effect of PISN mass regime shift on the observable fraction of PISNe. The (non-)detection of a non-super-luminous PISN could provide additional constraints on the stellar physics determining the PISN boundary.

The cut-off mass in the binary BH mass distribution could provide a similar constraint on the lower PISN boundary, except for other BBH formation channels polluting the PISN mass gap. However, since \textsc{bpass} already populates this regime through stable mass transfer and super-Eddington accretion with minimal dependence on the existence of PISN \citep{Briel+23}, we cannot probe the effect of a shifted PISN mass regime on the BBH population with \textsc{bpass}.

\subsection{Comparison to other PISN rate predictions}

Because many theoretical predictions for PISN rates and metallicity dependence are present in the literature, each with different assumptions for their rates \citep{langer_2007, Magg+16,Takahashi18, farmer_2019, duBuisson+20, Tanikawa+23}, we limit our comparison to population synthesis predictions and specific binary channels.
Our uniform metallicity prediction is similar to \citet{Tanikawa+23} and \citet{hendriks_2023}, who both use binary population synthesis for their predictions. The latter find rates between 3.6 and 60.5 PISN Gpc$^{-3}$ yr$^{-1}$ at z=0.028, dependent on the shift of PISN to higher or lower CO core masses, respectively.
The fiducial model with a maximum IMF mass of $300 \msun$ from \citet{Tanikawa+23}, which is most similar to our uniform model, predicts 1.2 PISN Gpc$^{-3}$ yr$^{-1}$ at $z=0$.
The peak of their model is at higher redshift, $z=6$, compared to our $z=3$ peak rate in the uniform-metallicity model. Moreover, our peak rate is 2 orders of magnitude smaller than theirs. This is most likely an effect of a fast enrichment of star formation in the IllustrisTNG simulation compared to the metallicity evolution of the \citet{madau_2017} prescription, despite \citet{Tanikawa+23} implementing stronger stellar winds than \textsc{bpass}.

\citet{duBuisson+20} estimate the pair-instability rate for chemically homogeneous evolving binaries resulting in one or more PISNe and their observability by relating $M_\mathrm{He, core}$ and peak magnitude of the supernova.
They predicted that HSC would have observed a few bright PISN already with a peak at $z\sim3$ based on an analytical description of the nickel produced during the PISN and if all binaries between $0.8<q<1$ evolve similarly to $q=1$ binaries. Because \textsc{bpass} does not track tides before initiating mass transfer in the stellar models and does not have $q=1$ stellar models, we cannot explore this formation channel. However, given a flat log distribution of mass ratio, the fraction of $q=1$ systems is minimal, although contact system might result in increasing this fraction.
The merger formation scenario for PISN makes up $\sim 11$ per cent of our total PISN rate (3.2 PISN Gpc$^{-3}$ yr$^{-1}$ at $z=0$) or $10^{-5}$ CCSN${}^{-1}$, similar to the conservative result found by \citet{vigna-gomez_2019}. 
In contrast to \citet{vigna-gomez_2019}, we find PISNe from a large collection of mergers between two main-sequence stars, and between main-sequence and post-main-sequence stars. Many of these stars still lose their hydrogen envelope before undergoing a PISN, except for metallicities below $10^{-5}$ where a few merger models explode as a RSG.

\citet{stevenson_2019} predict the smallest PISN rate of $< 10^{-2}$ PISN Gpc$^{-3}$ yr$^{-1}$ at $z=0$. While they do not consider the contribution of mergers or disrupted systems, the majority of our PISN come from primary stars, which are included in their rate. We attribute the difference to stronger stellar winds limiting their PISN metallicity bias function to below $Z\lesssim 0.002$, while ours extends to  $Z \lesssim 0.006$.

\subsection{Observational limits}

A confident observation of a PISN has not been made yet. Thus, using non-detection in the Pan-STARRS1 Medium Deep Survey, an upper limit on the intrinsic rate of $6 \times 10^{-6} R_\mathrm{CCSN}$ has been set by \citet{Nicholl+13}.
Theoretical studies find much higher intrinsic rates from their models than this upper limit and often compare their rate against the super-luminous supernova Type-I rate \citep[see for example][]{Briel+21, hendriks_2023}.
We would like to caution against such comparisons without careful consideration of the mass distribution of PISN progenitors.
First, while some PISNe are expected to be bright, we find that $75$ per cent of our intrinsic PISN rate below $z=4$ comes from the He70 and He80 light curve models. Around this mass range, Nickel production is limited and the supernova is predominantly powered by the explosion energy \citep{herzig_1990, Kasen+11, gilmer_2017}. These SNe would not exhibit typical PISN light curve features and could hide in the luminous supernova population with a brightness between the typical magnitudes of Type Ib/c SN and SLSN \citep{scannapieco_2005, gomez_2022}.
Secondly, the upper limit set by \citet{Nicholl+13} uses the $100 \msun$ from \citet{Kasen+11}, and thus, is only valid for super-luminous PISNe. The brightest PISNe in our model only contribute $\sim 3$ PISN Gpc$^{-3}$ yr$^{-1}$ at $z=0$ ($\sim 10^{-5} R_\mathrm{CCSN}$), which is less than the $10$ PISN Gpc$^{-3}$ yr$^{-1}$ ($10^{-5} R_\mathrm{CCSN}$) from \citet{Nicholl+13}. If the light curve and stellar modelling hold, super-luminous PISNe are a factor of 3 rarer than their low-mass counterparts at low redshift.

\begin{figure}
    \centering
    \includegraphics[width=0.45\textwidth]{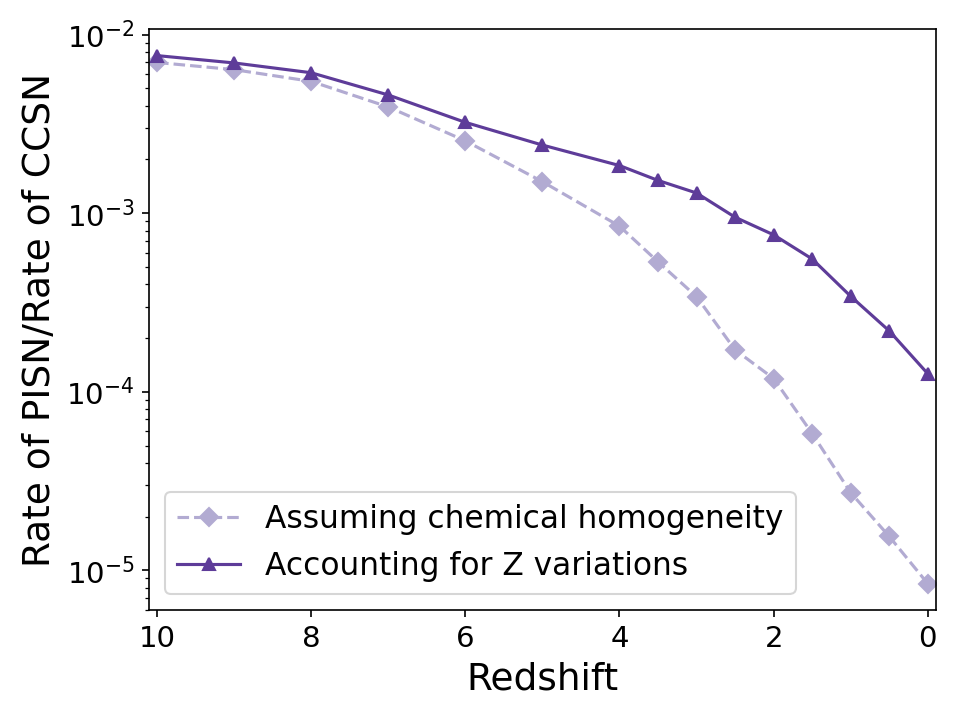}
    \caption{The PISN rate as a fraction of the CCSN rate over redshift for the uniform and detailed galaxy metallicity.}
    \label{fig:PISN/CCSN_rate}
\end{figure}

Furthermore, the PISN to CCSN fraction is non-uniform and decreased towards $z=0$ due to an increase in high metallicity star formation, as shown in Figure \ref{fig:PISN/CCSN_rate}. The CCSN rate tracks the star formation rate, mostly independent of metallicity, while the PISN rate still prefers low-metallicity star formation, which becomes scarcer at low redshift.

\subsection{Additional Caveats}

We matched the properties of our PISN progenitor to the properties of the models used for the light curves from \citet{Kasen+11}. However, the coverage of the possible parameter space in envelope and helium core masses is limited.
Figure \ref{fig:PISN_light_curve_tag} shows the tagging of light curves to the closest \citet{Kasen+11} light curve. Very few models are matched to the RSG light curves due to the \citet{Kasen+11} RSG models having a very large hydrogen envelope. Progenitor models with hydrogen envelope masses up to $\sim 38 \msun$ are tagged as a He model. The additional mass will lead to increased BH masses \citep{Winch+24} and will brighten the PISN, although most progenitors with a hydrogen envelope are in the already bright He PISN regime of 100 to 130 $\msun$ models. To refine our analysis, a larger suite of light curve models must be computed, for accurate characterisation of the large variety of expected visible PISN features. Such work would lead to more informative observational constraints, and inform light curve matching searches both with future surveys and in archival data.

Another essential aspect of these predictions is the predicted metallicity spread inside galaxies.
Compared to other cosmological simulations, the IllustrisTNG provides similar internal metallicity distributions \citep{ZZ2}, although they might be too wide compared to the Universe. A potential observed PISN rate might lie between the uniform and non-uniform predictions based on the amount of non-uniform metallicity in galaxies. However, due to limited observations of the spread of metallicities within galaxies across cosmic time, the cosmological simulations are the best predictions currently available for these properties.

\section{Conclusions} \label{sec:conclusions}

Although theoretically robust, the impact of internal stellar and binary physics on PISN can have drastic effects on their occurrences. In this work, we have explored the effect of accounting for non-uniform metallicity distributions in star-forming galaxies, as predicted by the IllustrisTNG simulation, on the cosmic PISN rate and the properties of their host galaxies. By linking the detailed stellar profiles from the binary population synthesis, \textsc{bpass}, to the PISN light curves from \citet{Kasen+11}, we have made detailed predictions for each their rate and the observability of PISN by the Euclid Deep Survey. These findings can be summarised as follows:

\begin{enumerate}
    \item The non-uniform metallicity distribution increases the cosmic PISN rate from $\sim2$ PISN Gpc$^{-3}$ yr$^{-1}$ at $z=0$ to 30 PISN Gpc$^{-3}$ yr$^{-1}$ at $z=0$.
    Based on the PISN light curves from \citep{Kasen+11}, we predict that the Euclid Deep Survey will observe 13.8 PISN per year (1.2 in the uniform metallicity case) for a total of 83 PISNe over the six-year lifetime of the Euclid mission.

    \item Only 12 per cent of the total intrinsic PISN rate at $z=0$ in our population results in a SLSN powered by Nickel decay. Our rate of bright PISN of 3.5 Gpc$^{-3}$ yr$^{-1}$ ($1.5 \times 10^{-5} R_\mathrm{CCSN}$ ) falls below the limit set by \citet{Nicholl+13}. Care should be taken when comparing intrinsic PISN rates to observed limits.
    
    \item We predict that 75 per cent of PISN have a 70 to 80 He star as a progenitor, which does not result in a bright SN and may only be sub-luminous \citep{herzig_1990, Kasen+11, gilmer_2017} and would not have been observable by the PanSTARRS1 Medium Deep Survey. Given our model assumptions and the supernova tagging, most of the PISNe would not have been classified as a SLSN-I nor SLSN-II. Observations of a non-super-luminous PISN could help us constrain the PISN mass range and, thus, stellar physics.

    \item PISN host galaxies have high star-formation rates, $\mathrm{SFR}_\mathrm{mean} \approx 0.1 \msun$ yr$^{-1}$, with a median galaxy mass of $10^9 \msun$ at $z=0$. While high star formation rates are preferred, the main driving factor for PISN probability is metallicity, especially compared to CCSN host galaxies. Even when considering a non-uniform metallicity distribution, the PISN occur in lower metallicity galaxies than the galaxies with the majority of star formation.
        
\end{enumerate}

More careful consideration of the PISN progenitors and their observability is essential in the comparison between theoretical PISN rates and observational limits. While the parameter space coverage of PISN light curves is still limited, many PISN might have avoided observation up to now.

While confident PISN observations have remained elusive, upcoming future deep surveys have a high likelihood of observing at least one bright PISN. With the launch of Euclid, the Euclid Deep Survey will provide valuable constraints on the PISN rate. 

\section*{acknowledgements}

\textit{Authors’ contributions:} 
MB and BM contributed equally to this work, and should be considered equally as co-lead authors for publication metrics.
MB led the use of the \textsc{bpass} models to compute metallicity biases for the various subcategories of PISN to examine their robustness, and was primarily responsible for writing the draft text and interpretation of observational rates.
BM proposed and led the implementation of the main modelling framework, using the IllustrisTNG simulations in order to predict the host galaxy properties and the number of visible PISNe from the metallicity biases input, and was primarily responsible for the definition and preparation of the figures in the paper.
TJM ran simulations to determine the discovery fractions of PISNe for HSC and Euclid.
JE provided expert support in the use of the \textsc{bpass} code and provided additional stellar models.
MT contributed advice on the design of the project and scientific objectives, data visualisation, interpretation, and structure of the manuscript.
All authors contributed comments during the research activities and the manuscript preparation.

The authors would like to thank Prof. Daniel Kasen for providing the theoretical PISN light curves, and the anonymous referee whose comments helped improve the quality of this work.

MMB is supported by the Boninchi Foundation and the Swiss National Science Foundation (project number CRSII5\_213497). MMB and JE acknowledge support from the University of Auckland and funding from the Royal Society Te Apar\={a}ngi of New Zealand Marsden Grant Scheme.

BM acknowledges support from an Australian Government Research Training Program (RTP) Scholarship. This research is supported in part by the Australian Research Council Centre of Excellence for All Sky Astrophysics in 3 Dimensions (ASTRO 3D), through project number CE170100013.

TJM is supported by the Grants-in-Aid for Scientific Research of the Japan Society for the Promotion of Science (JP24K00682, JP24H01824, JP21H04997, JP24H00002, JP24H00027, JP24K00668). TJM was supported by the Australian Research Council (ARC) through the ARC's Discovery Projects funding scheme (project DP240101786). TJM was supported by JSPS Core-to-Core Program (grant number: JPJSCCA20210003).

\section*{data availability}

\textsc{bpass} and its stellar models are available here: \url{bpass.auckland.ac.nz}.
We provide the data to create the metallicity bias functions here: \href{https://zenodo.org/records/13219564?token=eyJhbGciOiJIUzUxMiJ9.eyJpZCI6IjI3MTE5MTA5LWY2NjctNDFjZC04YTg0LTI3ZmM4NzhmMWYzNyIsImRhdGEiOnt9LCJyYW5kb20iOiIwNmJlNDg0Y2NiOGZkNTUzYzFmMWRiYjAwMGYzNTU1MCJ9.WzJB6GGHyFVjCbjqBDnJv9ZDf7_Ez4NvhQ0UBk3furCGkJDjrWC98R0bfM4scDLfqu7IkqTRiX10upV8NZ_HWg}{10.5281/zenodo.13219564}

\bibliographystyle{mnras}
\bibliography{biblio} 

\appendix

\section{PISN selection function}
\label{app:otherPISN_prescriptions}

\begin{figure}
    \centering
    \includegraphics[width=0.48\textwidth]{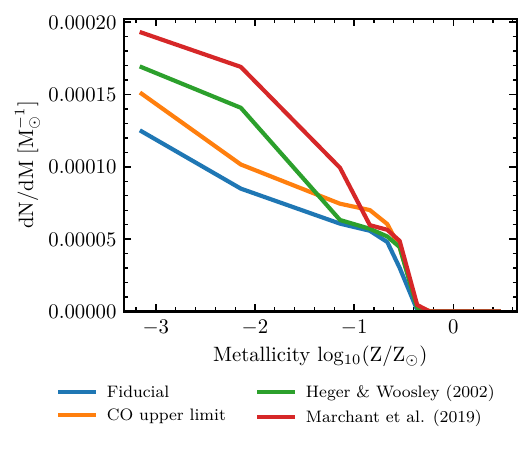}
    \caption{The fiducial metallicity bias function for PISN progenitors in this work (blue) compared against only a CO core lower limit (orange) and two other PISN prescriptions for non-rotating PISN progenitors. \citet{Heger+Woosley02} is the original \textsc{bpass} selection criteria, while \citet{marchant_2019} shifts the PISN regime to slightly lower masses, as described in the text.}
    \label{fig:PISN_bias_functions}
\end{figure}

In general, most PISN prescriptions use the He or CO core to determine if the progenitor star experiences pair instability and complete disruption. Figure \ref{fig:PISN_bias_functions} shows a collection of PISN prescriptions based on the He or CO core sizes. In the fiducial model, the CO and He core mass selection applied in this work are used. The $\mathrm{M}_\mathrm{CO}$ \textit{lower limit only} model contains all systems with $\mathrm{M}_\mathrm{CO} > 60 \msun$. The \citet{Heger+Woosley02} model is the original \textsc{bpass} prescription, where $64 \leq \mathrm{M}_\mathrm{He} < 133 \msun$, while the \citet{marchant_2019} model shifts this down to to  $60.8 \leq \mathrm{M}_\mathrm{He} < 124 \msun$.

The CO core selection criteria decrease the PISN metallicity bias function rate at most a factor of 1.6 at a metallicity of $10^{-4}$ (equivalent to $7.2 \times 10^{-3} Z_\odot$), but at other metallicities the impact is minimal. Because of this, the properties of PISN and their host galaxies discussed in this paper do not differ significantly when different PISN selection criteria are used.

\section{PISN formation channels}
\label{app:formation_pathways}

\begin{figure}
    \centering
    \includegraphics[width=0.49\textwidth]{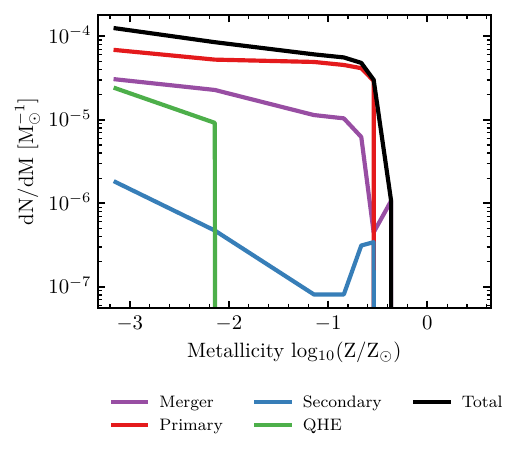}
    \caption{The formation pathways of PISN using the fiducial prescription. Left and right contain the same data, but are linear and log on the vertical axis, respectively. The Single star contribution is non-existent, because above $65 \msun$ no single stars are created in \textsc{bpass}, since above $10 \msun$ the multiplicity fraction is nearly 1 \citep{moe_2017}.} 
    \label{fig:formation_channels}
\end{figure}

Using the different stellar evolution pathways in \textsc{bpass}, we identify the progenitor model for each PISN. We plot the rate of PISNe coming from each channel as a function of their progenitor metallicities in Figure \ref{fig:formation_channels}. 
Here, \textit{primary} indicates that the primary star explodes as a PISN and is the dominant contribution to the PISN rate, which can occur with or without binary interaction.

When the companion instead explodes as a PISN, we denote it as a \textit{secondary}. These models have to retain or gain sufficient mass in previous evolutionary stages to explode as a PISN. Since the primary mass is by definition greater than the secondary mass in \textsc{bpass}, secondaries are less likely to end their lives as PISN. We note a slight increase in the contribution of the secondary channel around $Z = 3 \times 10^{-3}$ due to mass transfer.

The second largest contribution to the PISN rate comes from \textit{mergers}, where a main-sequence or post-main sequence star merges with a main-sequence companion. This increases the available hydrogen and allows for PISN up to higher metallicities. However, much of the hydrogen is still striped due to stellar winds, which are boosted in the host and luminous merger remnant.

A final contribution is quasi-homogenous evolution, \textit{QHE}, where due to accretion the secondary evolves in a chemically homogenous fashion. This only occurs if the companion accretes more than 5 per cent of its initial mass. Because it is only accretion that induces QHE, this contribution is limited and restricted to only the lowest of metallicities, where sufficient accretion and mass retention can occur. 

\section{PISN light curve progenitors}
\label{app:light_curves}

\begin{figure}
    \centering
    \includegraphics[width=0.49\textwidth]{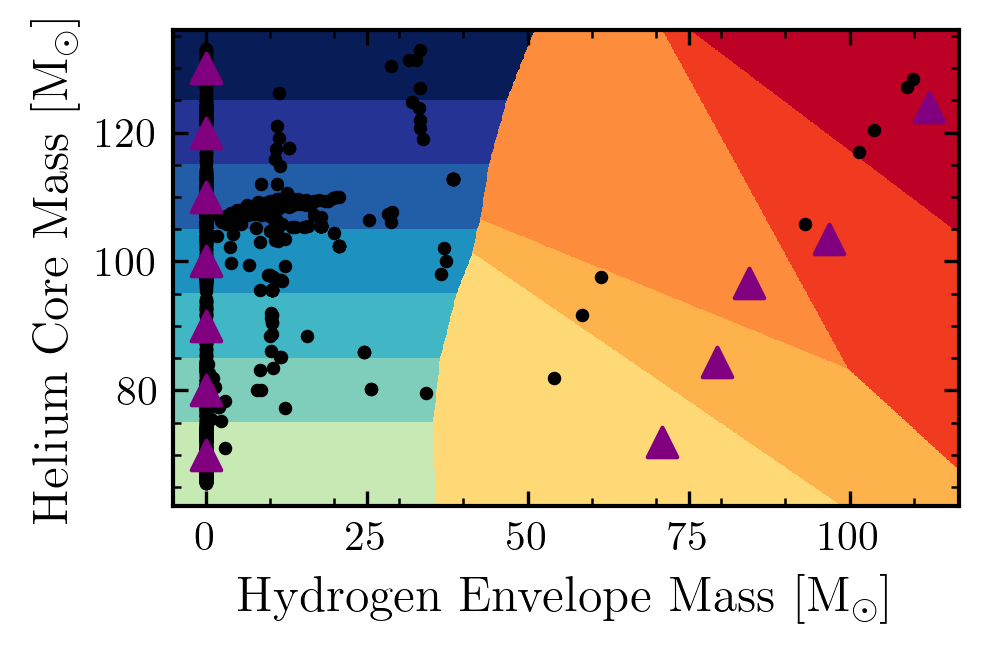}
    \caption{The classification of progenitor models parameters to the light curves of \citet{Kasen+11}. The purple triangles indicate the hydrogen envelope mass and helium core mass of the PISN progenitors models from \citet{Kasen+11}. Black points indicate each simulated PISN progenitor produced by BPASS at the end of its life. Each colour indicates the region where progenitor models are classified by a light curve model. The same colours are used as in Figure \ref{fig:PISN_lightcurve}.}
    \label{fig:PISN_light_curve_tag}
\end{figure}

To determine the observability of PISN by HSC and Euclid, we use the PISN light curves by \citet{Kasen+11}. We link the BPASS progenitor models to their PISN progenitor properties, by using a Voronoi tessellation. Considering the helium core mass and hydrogen envelope mass as the two most important properties that determine the light curve properties of a PISN, we partition the He-core/H-envelope plane by using the properties of each of our model light curves, and match each PISN simulated by BPASS to the closest PISN for which a light curve is available. We show this procedure graphically in Figure \ref{fig:PISN_light_curve_tag}. Here, we see that \textsc{bpass} progenitor models with a significant hydrogen envelope up to ${\sim} 30 \msun$ are tagged as helium progenitors. Correctly accounting for this additional material could drastically change their observability, since RSG light curves are generally brighter than the helium light curves. This does not affect the low mass regime, where all hydrogen is stripped from the PISN progenitor. 

\begin{figure}
    \centering
    \includegraphics[width=0.48\textwidth]{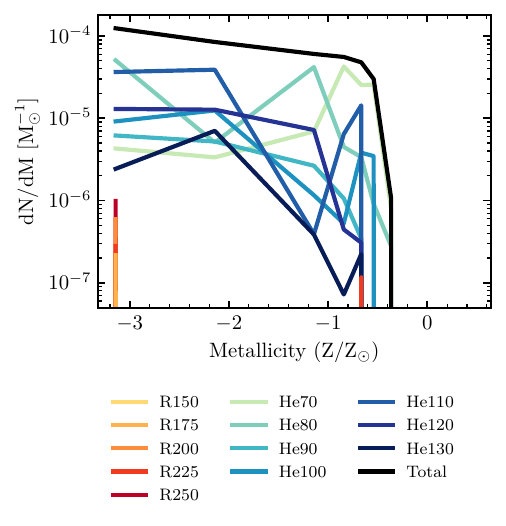}
    \caption{The PISN metallicity bias function per category of light-curve from \citet{Kasen+11}. No blue supergiants are present in the PISN progenitor population due to the minimal metallicity of \textsc{bpass}. The core and envelope masses are matched to the closest light-curve progenitor model. Red supergiant (R) models contribute only at the lowest metallicity, except for R225, which has a small  contribution at $Z=0.003$.}
    \label{fig:PISN_lightcurve}
\end{figure}

By summing the total rates of all PISN models associated with each lightcurve, we compute the metallicity bias function for each light curve, as shown in Figure \ref{fig:PISN_lightcurve}. RSG models contribute mostly at the lowest considered metallicity of $10^{-5}$, while all other progenitors are assigned a helium light curve model. We see the He70 and He80 models becoming dominant towards higher metallicity, while the more massive helium light curve models decrease, as expected, due to stronger stellar winds.

\begin{figure}
    \centering
    \includegraphics[width=0.5\textwidth]{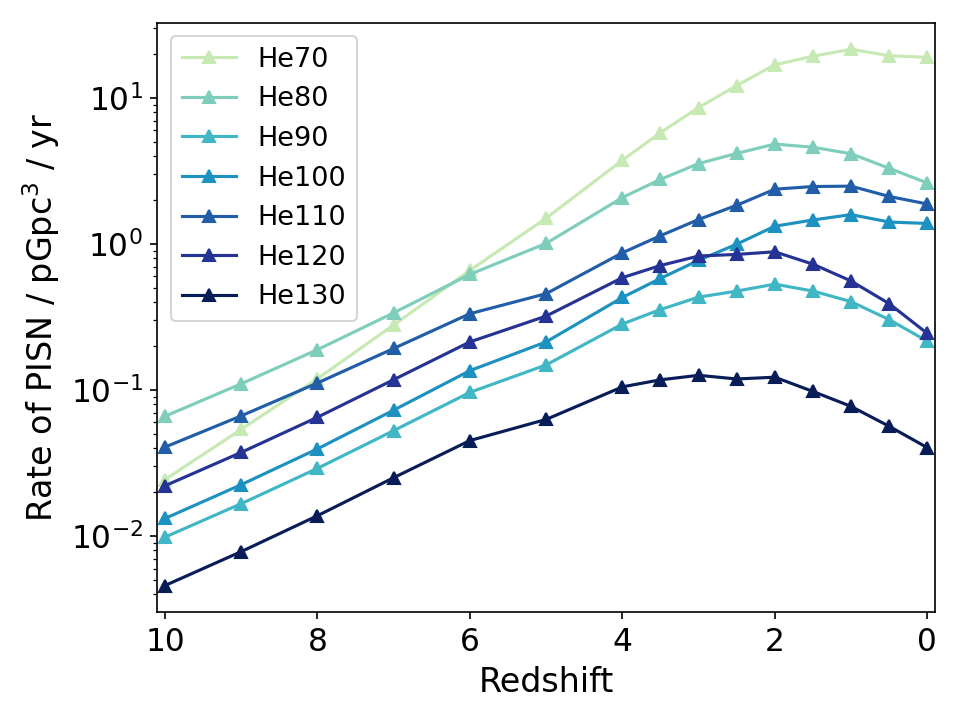}
    \caption{The rate per Gpc$^3$ per year for the helium light curves over redshift for the non-uniform metallicity distribution. He70 and He80 light curves make up the majority of the intrinsic PISN rate. Near $z=2$, star formation peaks, as do many of the light curves from massive helium progenitors.}
    \label{fig:He_lightcurves}
\end{figure}

\begin{figure}
    \centering
    \includegraphics[width=0.5\textwidth]{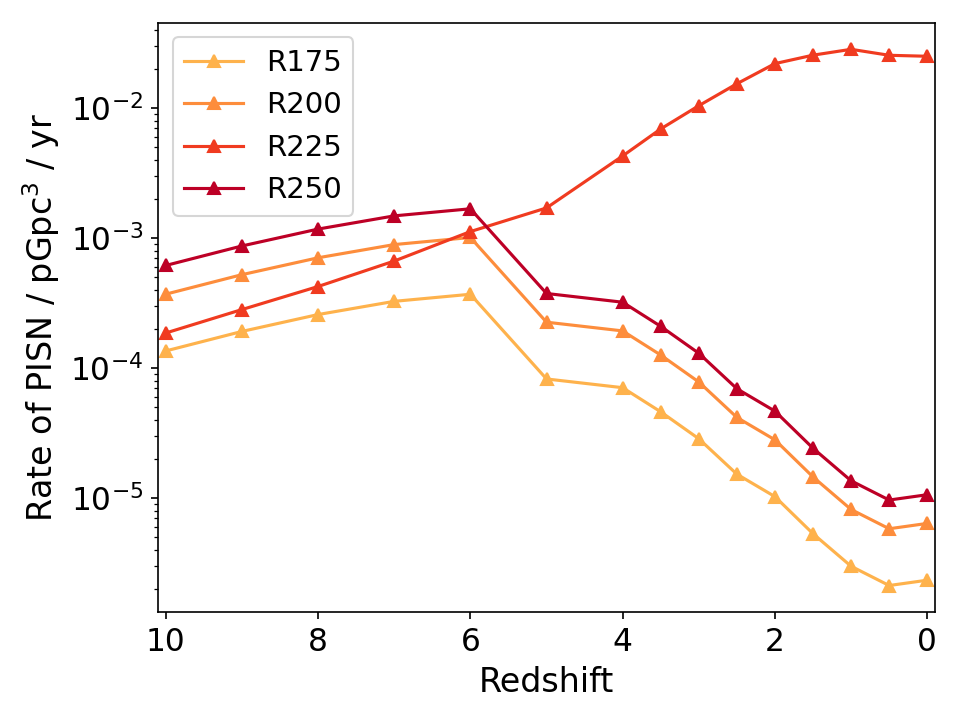}
    \caption{The rate per Gpc$^3$ per year for the RSG light-curves over redshift for the non-uniform metallicity distribution. Only a few models at low metallicity are tagged as these light curves, thus creating a decreasing trend with redshift. Except for R225, which has one model tagged at $Z=0.003$, which causes it to increase to lower redshifts. However, models with a hydrogen envelope are tagged as helium models due to the limited parameter space coverage of the light curve models.}
    \label{fig:RSG_lightcurves}
\end{figure}

Figures \ref{fig:He_lightcurves} and \ref{fig:RSG_lightcurves} show the predicted rate across cosmic time for all He-star models and all RSG models respectively. The RSG models contribution is sub-dominant due to the few hydrogen-rich progenitors and the limited coverage of envelope masses in light curve progenitor models, as we discussed in Section \ref{ssec:stellar_modelling}. The R225 model dominates this rate due to a small contribution at $Z \approx 0.21Z_\odot$, while the other models only occur at $Z \approx 7.2 Z_\odot$. However, the fraction of these models is minimal compared to the dominant He70 light curve. At $z=0$, the He70 channel makes up 65 per cent of the total PISN rate. Together, the low-mass He70 and He80 channels are the predicted final fates of 75 per cent of PISN progenitors. These light curves are expected to be faint and not produce a super-luminous supernova, thus, drastically reducing the likelihood of observing a PISN. Only 12 per cent of PISN have a helium core with a mass above $100 \msun$ and could potentially produce a super-luminous supernova.

\section{Detectable PISN fraction}
\label{app:detectable_fraction}

\begin{figure}
    \centering
    \includegraphics[width=0.5\textwidth]{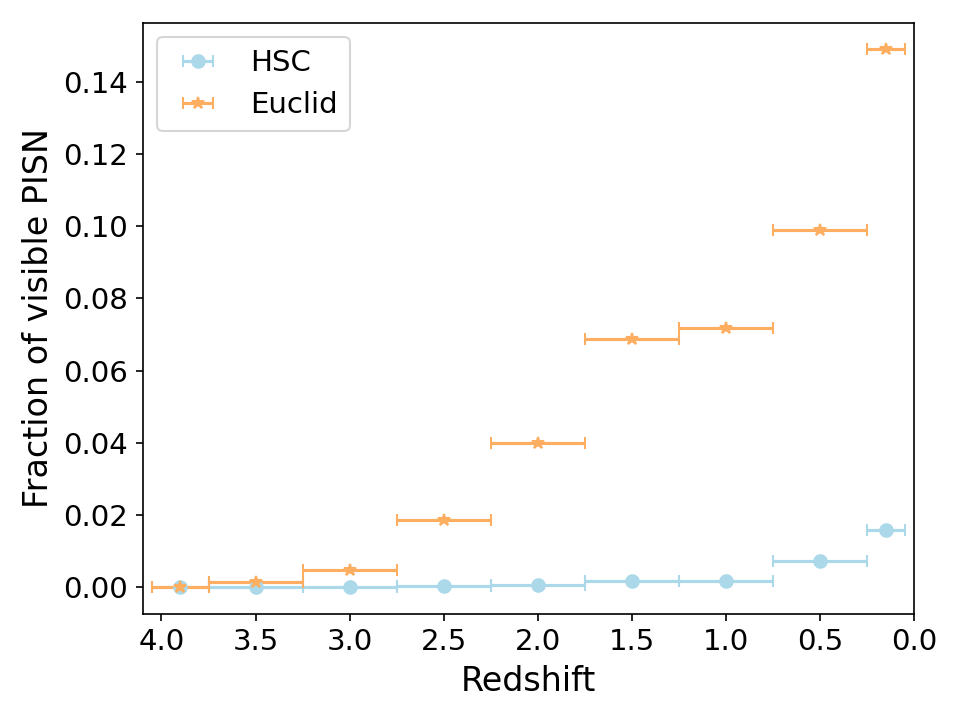}
    \caption{The fraction of PISN that are detectable by a typical HSC (blue circles) or the Euclid Deep Survey (orange triangles) as a function of redshift for the non-uniform metallicity model. Even in the local Universe, the vast majority of PISNe are expected to be too faint to be seen by these surveys.}
    \label{fig:vis_fracs}
\end{figure}

Based on the weight of each light curve model, we calculated the rate of observed PISNe in Figure \ref{fig:vis_rates}. Based on these rates and the intrinsic PISN rate, we calculate the visible fraction of PISN at each redshift in Figure \ref{fig:vis_fracs}. It shows that the HSC survey is very unlikely to observe any of the intrinsic PISNe, even in the local universe. EDS will be able to see ${\sim} 14$ per cent of all PISN at $z=0$. However, even its detection fraction decreases towards zero at $z=4$, making it difficult to probe the contribution of population III stars to the PISN, which is expected to become the dominant formation channel at $z=6$ \citep{Tanikawa+23}.

\label{lastpage}
\end{document}